\documentclass[english,prb,aps,amsmath,amssymb,reprint,superscriptaddress,showkeys]{revtex4-2}
\usepackage{amsmath}
\usepackage{mathptmx}
\usepackage{newtxmath}

\usepackage[T1]{fontenc}
\usepackage[utf8]{inputenc}
\usepackage{color}
\usepackage{babel}
\usepackage{graphicx}
\usepackage{wasysym}
\usepackage[unicode=true,pdfusetitle,
 bookmarks=true,bookmarksnumbered=false,bookmarksopen=false,
 breaklinks=true,pdfborder={0 0 0},pdfborderstyle={},backref=false,colorlinks=true]
 {hyperref}
\hypersetup{
 allcolors=blue}

\makeatletter

\providecommand{\tabularnewline}{\\}

\makeatother

\begin{document}
\title{Theory of shallow and deep boron defects in 4H-SiC}
\author{Vitor J. B. Torres}
\affiliation{I3N, Department of Physics, University of Aveiro, Campus Santiago,
3810-193 Aveiro, Portugal}
\author{Ivana Capan}
\affiliation{Ruđer Bošković Institute, Bijenička 54, 10000 Zagreb, Croatia}
\author{José Coutinho}
\email{jose.coutinho@ua.pt}

\affiliation{I3N, Department of Physics, University of Aveiro, Campus Santiago,
3810-193 Aveiro, Portugal}
\begin{abstract}
Despite advances toward improving the quality of $p$-type 4H-SiC
substrates and layers, we still have no model capable of accounting
for the multitude of boron-related optical, junction, and paramagnetic
resonance experiments available in the literature. A conspicuous puzzle
is the observation of two shallow boron defects with rather distinct
axial orientations as found by electron paramagnetic resonance (EPR)
and electron nuclear double resonance (ENDOR) data. This feature is
not observed in material doped with other group-III elements. Another
open issue involves conflicting conclusions from photoluminescence
and EPR studies of a deeper boron center, which has been linked to
rather distinct models, either based on substitutional or vacancy-related
boron defects. We unlock these and other problems by means of first-principles
calculations, where the temperature-dependent stability, the electronic
activity, and the paramagnetic response of boron defects in 4H-SiC
are investigated.\emph{ {[}Pre-print published in Physical Review
B }\textbf{\emph{106}}\emph{, 224112 (2022){]}}

\noindent \href{https://doi.org/10.1103/PhysRevB.106.224112}{DOI:10.1103/PhysRevB.106.224112}
\end{abstract}
\maketitle

\section{Introduction}

Due to is rugged properties, including mechanical, thermal, and chemical
stability, a large breakdown field, and the possibility of growing
both electronic-grade $n$- and $p$-type layers, 4H silicon carbide
(4H-SiC) is nowadays a semiconductor with an important and growing
market on power electronics (used in electric vehicles, power supplies,
motor control circuits, and inverters) \citep{Kimoto2014,Liu2015}.
SiC also finds applications in fundamental and emerging fields like
high-energy particle detection \citep{Coutinho2021} and quantum technologies
\citep{Lukin2019,Castelletto2020,Wolfowicz2021,Anderson2022}.

The $p$-type dopants are usually boron, aluminum, and gallium. As
for the former, there is ample evidence that its incorporation leads
to the appearance of two types of acceptors, often referred to as
\emph{shallow} and \emph{deep} boron centers, owed to the relative
depth of their respective levels within the band gap \citep{Suttrop1990,Sridhara1998}.
The two boron species diffuse differently --- boron-implanted/diffused
layers show heterogeneous incorporation, where the deep center dominates
the profile tails \citep{Gao2003,Bockstedte2004,Aleksandrov2013}.
While the assignment of the shallow species to substitutional boron
on the Si site ($\textrm{B}_{\textrm{Si}}$) seems consensual, the
origin of the deep hole trap has remained elusive. Photoluminescence
studies favor a boron atom on the carbon site ($\textrm{B}_{\textrm{C}}$)
\citep{Kuwabara1975}, magnetic resonance experiments point to a boron-vacancy
complex \citep{Baranov1998,DuijnArnold1998}, whereas first-principles
results suggest either $\textrm{B}_{\textrm{C}}$ \citep{Bockstedte2001,Bockstedte2004}
or a boron-silicon-antisite pair \citep{Aradi2001}.

Another problem is that boron is often present in the SiC as a contaminant
in trace concentrations. The deep species, also referred to as D-center,
is of particular concern, especially in $n$-type SiC where it is
negatively charged under equilibrium conditions. This state is a potential
trap for holes, threatening the functioning of bipolar devices or
$n$-type detectors \citep{Storasta2002}.

A possible route for the elimination of the D-center involves thermal
oxidation \citep{Kawahara2013,Okuda2015,Ayedh2017}. However, the
impact of boron-related minority carrier lifetime degradation is not
necessarily detrimental. The effect was actually explored to improve
the switching time characteristics of $p$-i-$n$ diodes, and that
was attributed to the effect of a localized lifetime control in the
intrinsic layer due to carrier recombination at deep boron traps \citep{Bolotnikov2007,Yang2019}.

The D-center is known since early deep level transient spectroscopy
(DLTS) studies of B doped 6H-SiC, where two nearly overlapping peaks
corresponding to electronic transitions at $E_{\textrm{v}}+0.63$~eV
and $E_{\textrm{v}}+0.73$~eV were revealed \citep{Anikin1985}.
Suttrop \emph{et~al.} \citep{Suttrop1990} found that in addition
to the deep boron center (measured in that work as a single DLTS peak
at $E_{\textrm{v}}+0.58$~eV), a hole trap at $E_{\textrm{v}}+0.30$~eV
was also present, and it was assigned to the shallower boron acceptor.

The presence of the D-center in the 4H polytype was also confirmed
using DLTS by Sridhara \emph{et~al.} \citep{Sridhara1998}. The level
was placed at $E_{\textrm{v}}+0.55$~eV (assuming a $T^{-2}$-corrected
cross-section), again without resolving a double peak structure. Although
the shallower species could not be found by DLTS (the Si/C ratio of
the samples did not favor its formation), admittance spectroscopy
measurements of Si-poor samples arrived at an acceptor level for shallow
boron in the range 284-295~meV above $E_{\textrm{v}}$ \citep{Sridhara1998}.

Recently, Laplace-DLTS and Laplace-minority carrier transient spectroscopy
(Laplace-MCTS) measurements were carried out for studying the shallow
and deep boron centers in 4H-SiC \citep{Capan2020}. Estimated activation
energies for hole emission were respectively 0.27~eV and 0.60~eV.
From Laplace-MCTS, it was shown that the D-center consists of two
components, D1 and D2 with nearly 1:1 intensity ratio, respectively
estimated at $E_{\textrm{v}}+0.49$~eV and $E_{\textrm{v}}+0.57$~eV.
The pair of traps was assigned to boron at two different carbon sublattice
locations in 4H-SiC. The peak of the shallow boron species was structureless.
If it corresponded to the superposition of more than one point defect
(in different sublattice sites), they were indistinguishable as far
as the resolution offered by Laplace-MCTS.

Early electron paramagnetic resonance (EPR) studies \citep{Zubatov1985}
indicated that the symmetry of the shallow boron species in 6H-SiC
experienced a remarkable change upon lowering the temperature. In
the 6H phase, two cubic ($k_{1}$ and $k_{2}$) and one hexagonal
($h$) sites are available for B$_{\textrm{Si}}$ substitution. While
above $T=50$~K the EPR signals related to all tree substitutions
show a trigonal pattern, below that temperature the $k$-related signals
lower their symmetry to monoclinic. The $h$-related signal preserves
$C_{3v}$ symmetry for temperatures as low as 5~K.

These findings were confirmed latter by electron nuclear double resonance
(ENDOR) spectroscopy \citep{Muller1993,Matsumoto1997}. The defect
structure was interpreted as comprising a B-C broken bond, where boron
is threefold coordinated (connected to three C ligands), while the
remaining C atom holds a hole that is responsible for 40\% of the
total spin density. Strikingly, whereas the C radical is aligned along
the main crystallographic axes for the case of $\textrm{B}_{\textrm{Si}}$
sitting on the $h$ site, for some reason, boron on the cubic sites
leave a C dangling bond aligned along a basal B-C direction. Analogous
observations were reported in 4H-SiC samples \citep{GreulichWeber1997,GreulichWeber1998}.

The deep boron center also has EPR-related signals and several experiments
produced rich amounts of data (see Refs.~\onlinecite{DuijnArnold1998},
\onlinecite{Baranov1998} and references therein). The defect has
a spin-1/2 paramagnetic state, but unlike the shallow boron center,
both $h$- and $k$-related signals show the same alignment along
the hexagonal $c$ axis, with a small basal anisotropy. Minute $^{13}\textrm{C}$
satellite lines were detected around the main signals, the $^{11}\textrm{B}$
hyperfine interactions were negligible, and no large $^{29}\textrm{Si}$
were observed either. However, the spin density was found to be almost
100\% localized on Si ligands. Based on the data, a model combining
a boron on a silicon position with an adjacent carbon vacancy ($\textrm{B}_{\textrm{Si}}\textrm{-}V_{\textrm{C}}$)
was proposed. The structure comprises an inert boron atom and three
Si radicals edging the $V_{\textrm{C}}$ unit, thus explaining the
electronic and magnetic activity \citep{DuijnArnold1998,Baranov1998}.
An obvious difficulty of this model is that for some reason, the pair
would have to be invariably formed with an alignment along the $c$
axis. Such preferential alignment is not supported by first-principles
modeling. In fact, the calculations also show that the lowest-lying
level of $\textrm{B}_{\textrm{Si}}\textrm{-}V_{\textrm{C}}$ is a
donor in the upper half of the gap, and therefore, the complex is
not compatible with the D-center \citep{Aradi2001,Bockstedte2001}.

While early semi-empirical Hartee-Fock calculations using small H-terminated
SiC clusters predicted that $\textrm{B}_{\textrm{Si}}$ adopts an
off-center configuration \citep{Bratus1993,Petrenko1996}, subsequent
supercell calculations within the local density approximation (LDA)
to density functional theory (DFT) led to ambiguous conclusions. Accordingly,
some authors justified the off-site location of $\textrm{B}_{\textrm{Si}}$
with a Jahn-Teller (JT) effect \citep{Fukumoto1996,Bockstedte2001}.
Others found an effective-mass-like defect with no distortion at all
\citep{Deak2003}. Finally, the authors of Ref.~\onlinecite{Gerstmann2004}
found that the LDA cannot describe the shallow boron state due to
overmixing with the valence band. After applying a scissors correction
to the band gap during the self-consistent Kohn-Sham method, they
obtained a pronounced JT distortion toward $C_{1h}$-symmetry and
a prominent $^{13}\textrm{C}$-hyperfine interaction due to a C-radical
\citep{Gerstmann2004}. Although they account for the measured localization
of the spin density, these results cannot explain the different symmetries
of $k$- and $h$-related boron EPR signals in both 4H- and 6H-SiC.

It is well-known that local and semilocal approximated DFT poorly
describes insulator/semiconductor band gaps, making the discussion
of defect properties, in particular those that involve gap states,
vulnerable. For instance, several insufficiencies of conventional
DFT and advancements in modeling the electronic structure of defects
in SiC were presented in Ref.~\citep{Oda2013}. Among the findings
it was shown that hybrid DFT, which replaces a fraction of the local
exchange potential by a (possibly screened) Fock exchange contribution,
can provide reliable electronic structure of defects in SiC where
a local density description fails. We revisited the theory of substitutional
boron defects to verify if modern electronic structure calculation
methods, in particular hybrid density functional theory, can shed
light on the open issues described above. After detailing the methods
employed in Sec.~\ref{sec:theory}, we report on the physical picture
of Si and C replacements by boron in 4H-SiC (Secs.~\ref{subsec:sb}
and \ref{subsec:db}). The following three sections connect our findings
with photoluminescence and junction capacitance spectroscopies (Sec.~\ref{subsec:dlts}),
with finite-temperature effects on the preferential formation of Si
or C substitutions (Sec.~\ref{subsec:ft}), as well as with the available
EPR/ENDOR measurements (Sec.~\ref{subsec:epr}).

We show that $\textrm{B}_{\textrm{Si}}$ and $\textrm{B}_{\textrm{C}}$
defects nicely explain the optical, capacitance and magnetic measurements
related to shallow and deep boron centers in 4H-SiC, respectively.
Importantly, it is argued that the \emph{shallow} label attributed
to $\textrm{B}_{\textrm{Si}}$ should be interpreted as \emph{shallower}
than the deep boron center. In other words, the $\textrm{B}_{\textrm{Si}}$
center has the characteristics of a localized and deep hole trap and
not of an effective mass theory (EMT) dopant. The EMT picture for
$\textrm{B}_{\textrm{Si}}$ has been advocated based on (semi-)local
density functional results, but we show that higher level hybrid DFT
predicts a strong atomistic relaxation upon hole capture at a $\sim\!0.3$~eV
deep trap, making the model compatible with the magnetic resonance
observations. We rule out an assignment of deep boron to $\textrm{B}_{\textrm{Si}}\textrm{-}V_{\textrm{C}}$
based on the calculated $g$ tensor elements. Along the paper, we
also solve several problems, most notably we explain the observation
of different orientations of $g$ tensor and hyperfine interactions
for shallow boron on cubic and hexagonal sites and the distinct temperature-dependence
of the $g$ tensors of both centers.

\section{Theoretical Methods\label{sec:theory}}

First-principles calculations were carried out using the density functional
Vienna \emph{ab initio} simulation package (VASP) \citep{Kresse1993,Kresse1994,Kresse1996a,Kresse1996b},
employing the projector-augmented wave method, thus avoiding explicit
treatment of core states \citep{Blochl1994}. A basis set of plane-waves
with kinetic energy of up to 400~eV was used to describe the Kohn-Sham
states. Total energies were evaluated self-consistently, using the
hybrid density functional of Heyd-Scuseria-Ernzerhof (HSE06) \citep{Heyd2003,Krukau2006}
with a numerical accuracy of $10^{-7}$ eV. When compared to generalized
gradient approximated (GGA) calculations \citep{Perdew1996} ---
which underestimate the band gap of SiC by a factor of nearly one
half --- the HSE06 functional has the main advantage of predicting
a Kohn-Sham band gap width of 3.17~eV for 4H-SiC. This figure should
be compared to the experimental value of 3.27 eV \citep{Grivickas2007}.

Defect energies were found using 400-atom (defect-free) supercells
of 4H-SiC (with hexagonal shape), obtained by replication of $5\!\times\!5\!\times\!2$
primitive cells, into which boron defects were inserted. The equilibrium
(calculated) lattice parameters of 4H-SiC were $a=3.071$~Å and $c=10.052$~Å.
These are close to the experimental values of $a=3.079$~Å and $c=10.081$~Å
\citep{Stockmeier2009}.

Defect structures were firstly optimized within the HSE06 approximation
using $\mathbf{k}=\Gamma$ to sample the Brillouin zone (BZ), until
the largest force became lower than 0.01~eV/Å. On a second step,
electronic total energies of the obtained structures were found from
single-point calculations with the band structure sampled at a $\Gamma$-centered
$2\!\times\!2\!\times\!2$ mesh of $\mathbf{k}$-points (also within
HSE06). In line with Gerstmann \emph{et~al.} \citep{Gerstmann2004},
we found that structural optimizations of B$_{\textrm{Si}}$ defects
within the GGA led to erroneous results due to overmixing of gap states
with the valence band top. An analogous effect attributed to the overmixing
of a carbon interstitial ($\textrm{C}_{\textrm{i}}$) level, in this
case with the SiC conduction band bottom, was also pointed out by
Gouveia and Coutinho \citep{Gouveia2019}, and that will be further
discussed below.

Electronic transitions of boron defects were calculated by finding
the Fermi energy at crossing points of formation energies for different
charge states $q$. Defect formation energies ($E_{\textrm{f}}$)
were obtained as a function of the chemical potential of the ``sample''
constituents, according to the usual formalism (see for instance Refs.~\onlinecite{Qian1988}
and \onlinecite{Coutinho2020}),

\begin{equation}
E_{\textrm{f}}(\mu_{i},\mu_{\textrm{e}};q)=E_{\textrm{elec}}(q)-\sum_{i}n_{i}\mu_{i}-n_{\textrm{e}}\mu_{\textrm{e}}.\label{eq:eform}
\end{equation}
The first term on the right-hand side of Eq.~\ref{eq:eform} is given
by $E_{\textrm{elec}}(q)=\tilde{E}_{\textrm{elec}}(q)+E_{\textrm{corr}}(q)$,
and refers to the electronic energy of the periodic calculation $\tilde{E}_{\textrm{elec}}$
shifted by $E_{\textrm{corr}}$ to remove the effect of the artificial
and infinite array of localized charges when the charge state is $q\neq0$.
For that we use the method proposed by Freysoldt, Neugebauer, and
Van de Walle \citep{Freysoldt2009}, generalized for anisotropic materials
by Kumagai and Oba \citep{Kumagai2014}. The method uses the axial
and transverse dielectric constants of 4H-SiC, calculated as $\epsilon^{\ensuremath{\parallel}}=10.65$
and $\epsilon^{\ensuremath{\bot}}=9.88$, respectively \citep{Coutinho2017}.
See Ref.~\citep{Supplemental} (and also, Refs.~\citep{Makov1995,Castleton2006,Freysoldt2009,Lany2008,Kumagai2014,Coutinho2017})
for convergence tests to the formation energy of boron defects upon
varying the boundary conditions. The second and third terms sum up
the chemical potentials $\mu_{i}$ of the $n_{i}$ neutral atomic
species and $n_{\textrm{e}}=-q$ extra electrons (with respect to
the neutral state) that form the problem. The electronic chemical
potential is $\mu_{\textrm{e}}=E_{\textrm{v}}+E_{\textrm{F}}$, where
$E_{\textrm{v}}$ and $E_{\textrm{F}}$ are the valence band top and
Fermi energies, respectively. The former is obtained as the highest
occupied state in a bulk supercell, whereas the latter is an independent
variable.

Chemical potentials for $i=\{\textrm{Si},\textrm{C}\}$ were calculated
as

\begin{equation}
\mu_{i}=\mu_{i}^{0}+(1-f_{i})\Delta E_{\textrm{SiC}}^{\textrm{f}},\label{eq:mu}
\end{equation}
where $\mu_{i}^{0}$ are energies per atom in pure silicon or carbon
(diamond phase), $\Delta E_{\textrm{SiC}}^{\textrm{f}}$ is the heat
of formation of SiC estimated as $\Delta E_{\textrm{SiC}}^{\textrm{f}}=\mu_{\textrm{SiC}}^{0}-\mu_{\textrm{Si}}^{0}-\mu_{\textrm{C}}^{0}=-0.62$~eV,
with $\mu_{\textrm{SiC}}^{0}$ being the energy per SiC formula unit
in a perfect 4H-SiC crystal. This result is close to the enthalpy
of formation $\Delta H_{\textrm{SiC}}^{\textrm{f}}=-0.72$~eV measured
at standard conditions \citep{Greenberg1970}. Eq.~\ref{eq:mu} allows
for a variation of the chemical potentials in the range $\mu_{i}^{0}+\Delta E_{\textrm{SiC}}^{\textrm{f}}\leq\mu_{i}\leq\mu_{i}^{0}$
subject to $0\leq f_{i}\leq\sum_{i}f_{i}=1$, with the upper limit
representing $i$-rich conditions during the material growth. We will
calculate the relative energy of different boron defects, all of which
possessing a single boron atom. Although being an irrelevant quantity
for this purpose, the chemical potential of boron ($\mu_{\textrm{B}}$)
was found from the $\alpha$-rhombohedral ground state phase {[}12
atoms per unit cell with $R\bar{3}m$ space group (group No. 166){]},
with equilibrium lattice parameters $a=5.029$~Å and $\alpha=58^{\circ}$
\citep{Widom2008}.

We also examined the relative stability of boron acceptors on different
lattice sites at finite temperatures. The range of temperatures close
to those experienced during epitaxial growth are of particular importance.
In this case intrinsic conditions apply, and for acceptors with levels
in the lower part of the gap the relevant charge state is the negative
one. The difference in the Helmholtz free energy of formation between
two boron dopants replacing different crystalline species is obtained
as,

\begin{equation}
F(\textrm{B}_{\textrm{Si}}^{-})-F(\textrm{B}_{\textrm{C}}^{-})=\Delta F_{\textrm{elec}}+\Delta F_{\textrm{vib}}+\mu_{\textrm{Si}}-\mu_{\textrm{C}}.\label{eq:dfree}
\end{equation}

In the above, $\Delta F_{\textrm{elec}}=F_{\textrm{elec}}(\textrm{B}_{\textrm{Si}}^{-})-F_{\textrm{elec}}(\textrm{B}_{\textrm{C}}^{-})$
is the electronic free energy difference between the two defects,
where $F_{\textrm{elec}}$ is replaced by the stationary solution
of the electronic problem, $E_{\textrm{elec}}$ (obtained within hybrid
density functional theory). This approximation essentially neglects
electronic entropy, and it is justified by the depth of the electronic
levels and the negligible density of defect states at the Fermi level
under intrinsic conditions \citep{Estreicher2004}. The second term
on the right-hand side of Eq.~\ref{eq:dfree} accounts for the vibrational
free energy difference between the defects, and for each we have,

\begin{equation}
F_{\textrm{vib}}(T)=k_{\textrm{B}}T\sum_{i=1}^{3N-3}\ln\left[2\sinh\left(\frac{\hslash\omega_{i}}{2k_{\textrm{B}}T}\right)\right].\label{eq:fvib}
\end{equation}

The summation above runs over $3N-3$ vibrational modes of the $N$-atom
defective supercell, with respective angular frequencies $\omega_{i}$.
Symbols $k_{\textrm{B}}$ and $\hslash$ refer to the Boltzmann and
reduced Planck constants, respectively. It is noted that Eq.~\ref{eq:fvib}
already accounts for zero-point motion. Chemical potentials in Eq.~\ref{eq:dfree}
were found from Eq.~\ref{eq:mu}, after adding a vibrational term
$F_{\textrm{vib}}/N$ to $\mu_{\textrm{Si}}^{0}$ and $\mu_{\textrm{C}}^{0}$,
obtained from respective supercells of silicon and diamond made of
$N=64$ atoms, and with the temperature set to $T=273.15$~K. Analogously,
a vibrational free energy term $2F_{\textrm{vib}}/N$ was added to
$\mu_{\textrm{SiC}}^{0}$For further details regarding the calculation
of defect free energies, we direct the reader to Refs.~\onlinecite{Estreicher2004},
\onlinecite{Murali2015}, \onlinecite{Gomes2022} and references therein.

The vibrational mode frequencies of 4H-SiC cells containing boron
defects were evaluated in $N=72$-atom cells ($3\!\times\!3\!\times\!1$
primitive cells). We considered the participation of all atoms in
the dynamical matrix, whose elements were found from the force derivatives
with respect to the atomic positions \citep{Gomes2022}.

The $g$ tensor and hyperfine (HF) interactions of paramagnetic boron
defects were calculated using the gauge including projector augmented
wave (GIPAW) method \citep{Pickard2002} as implemented in the QUANTUM
ESPRESSO package \citep{Giannozzi2009,Giannozzi2017}. The GIPAW method
is based on self-consistent density functional perturbation theory,
describing the applied magnetic field and spin-orbit couplings as
perturbations. The current implementation pertaining the $g$ tensor
calculation is limited to local and semilocal functionals. Hence,
for these calculations, the Kohn-Sham states were found within the
GGA \citep{Perdew1996}. We used hexagonal supercells of 256 atoms,
a $\Gamma$-centered BZ sampling mesh of $2\!\times\!2\!\times\!2$,
and a plane-wave cutoff $E_{\textrm{cut}}=612$~eV (45~Ry). The
computation of reciprocal space derivatives to obtain spin currents
in linear magnetic response, makes the calculation of $g$ tensors
rather sensitive to $\mathbf{k}$-point sampling \citep{Pickard2001}.
Convergence issues can be especially severe for states whose $g$
values show large deviations from that of the free-electron. For that
reason, we also tested a denser $3\times3\times3$ grid in the evaluation
of $g$ values for neutral B$_{\textrm{Si}}$.

Due to erroneous geometries obtained for B$_{\textrm{Si}}$ defects
within GGA, atomistic structures for the GIPAW calculations were found
within HSE06-level (using the VASP code). Such combined approach was
successfully used in a recent study of defects in $\textrm{Ga}_{2}\textrm{O}_{3}$
\citep{Skachkov2019}.

As for the HF coupling tensors $A$, they describe the interaction
between the electron spin of a paramagnetic state with magnetic nuclei
at the defect core. For an axial state along an arbitrary principal
direction 3, transverse principal values $A_{1}=A_{2}$ and the HF
tensor can be described by isotropic ($a$) and anisotropic ($b$)
hyperfine constants, which relate to the diagonalized tensor components
as $a=(2A_{1}+A_{3})/3$ and $b=(A_{3}-A_{1})/3$ \citep{Muller1993}.
The evaluation of the HF tensors rely on the accurate computation
of the spin density embedding the nuclei of interest, and for the
case of the isotropic term (also known as Fermi contact), it involves
the description of the electron density within the core region. Therefore
the use of pseudopotentials implies a core reconstruction from the
pseudo-wavefunctions \citep{VandeWalle1993}.

\section{Results\label{sec:results}}

\subsection{Boron on the silicon site: shallow boron\label{subsec:sb}}

We start by looking at the boron impurity on the Si site. In the neutral
charge state, the boron atom was clearly displaced from the perfect
lattice site after optimizing the energy with respect to the atomistic
geometry. Essentially, boron formed three B-C bonds, leaving an unsaturated
C radical. The on-site structure was metastable with a small $\sim0.1$~eV
barrier along the way toward the off-site ground state structure.

\noindent 
\begin{figure}
\includegraphics[clip,width=7.5cm]{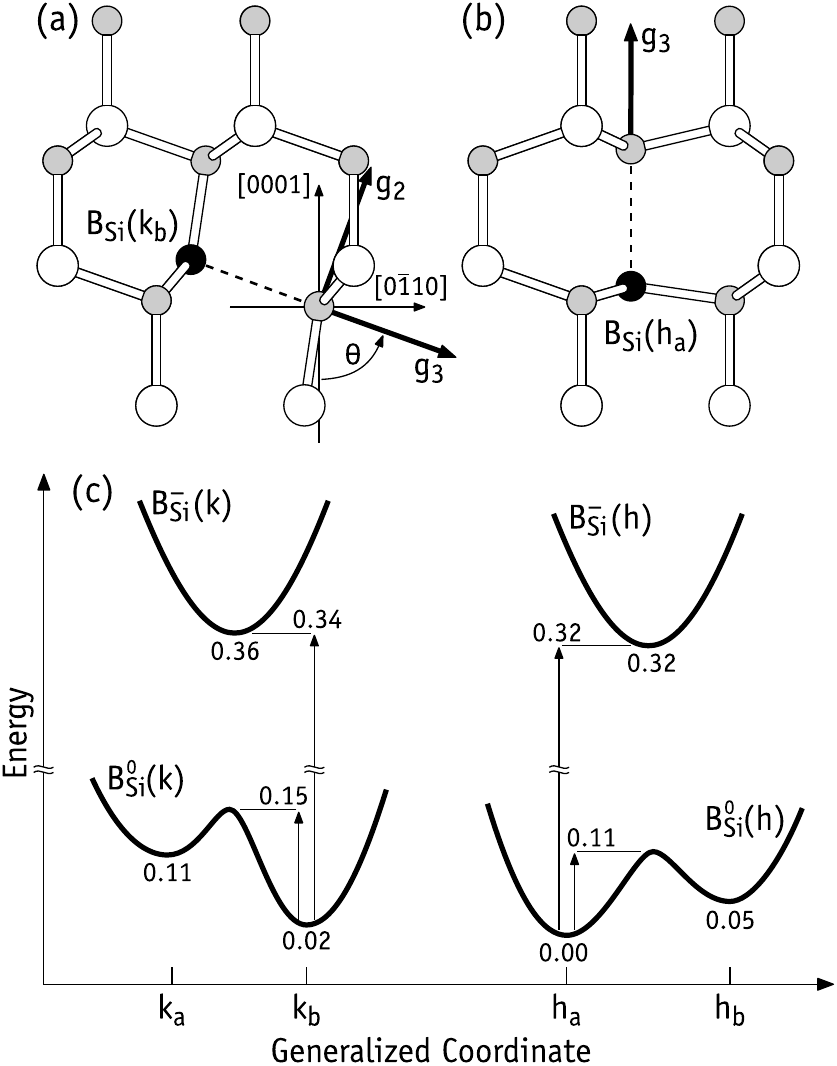}

\caption{\label{fig1}Low energy structures of neutral $\textrm{B}_{\textrm{Si}}$
at (a) $k$ and (b) $h$ sites of 4H-SiC, respectively, and configurational
coordinate diagram of neutral and negatively charge states (c). $g$
tensor principal directions of neutral states are also shown. Boron,
carbon and silicon are shown in black, gray and white, respectively.
All energies in the diagram are in eV. Energies located below the
energy minima are relative the $\textrm{B}_{\textrm{Si}}^{0}(h_{\textrm{a}})$
ground state. Energies next to arrow heads are relative to the state
next to the arrow base. See Ref.~\citep{Supplemental} for details
regarding the barrier calculations.}
\end{figure}

4H-SiC has two distinct sublattice sites, namely cubic ($k$) and
hexagonal ($h$), and for each, substitutional boron atoms can form
two types of C radicals, namely those polarized along the hexagonal
axis of the crystal (labeled with ‘a’ and standing for ‘axial’) and
those polarized along the basal bond directions (labeled with ‘b’
and standing for ‘basal’). This leads to a total of four possible
defect configurations to consider.

Among all structures, those depicted in Figs.~\ref{fig1}(a) and
\ref{fig1}(b), namely $\textrm{B}_{\textrm{Si}}(k_{\textrm{b}})$
and $\textrm{B}_{\textrm{Si}}(h_{\textrm{a}})$, were the most stable
at $k$ and $h$ sites, respectively. The B atom in both structures
displays threefold coordination, where three short (1.65~Å) B-C bonds
contrast with the $\sim\!2.42$~Å long separation between B and the
C radical (see dashed lines in Fig.~\ref{fig1}). See Ref.~\citep{Supplemental}
(and also Ref.~\citep{Monkhorst1976}), which provides further geometrical
details of the structures. While B-C bond lengths are essentially
the same for all configurations, the longer B-C distance can vary
by about 0.04~Å, depending on the specific site and orientation.
The energies of the two most stable neutral states, namely $\textrm{B}_{\textrm{Si}}^{0}(k_{\textrm{b}})$
and $\textrm{B}_{\textrm{Si}}^{0}(h_{\textrm{a}})$, differ by 0.02
eV only, whereas $\textrm{B}_{\textrm{Si}}^{0}(k_{\textrm{a}})$ and
$\textrm{B}_{\textrm{Si}}^{0}(h_{\textrm{b}})$ are metastable, respectively
at 0.11~eV and 0.05~eV above the ground state $\textrm{B}_{\textrm{Si}}^{0}(h_{\textrm{a}})$.
The reason for the breaking of the B-C bond along different directions
for $\textrm{B}_{\textrm{Si}}(k)$ and $\textrm{B}_{\textrm{Si}}(h)$
will become evident when we discuss the electronic structure of the
center further below. We summarize the above results in the lower
part of the configurational coordinate diagram represented in Fig.~\ref{fig1}(c).

\noindent 
\begin{figure*}
\includegraphics[clip,width=17cm]{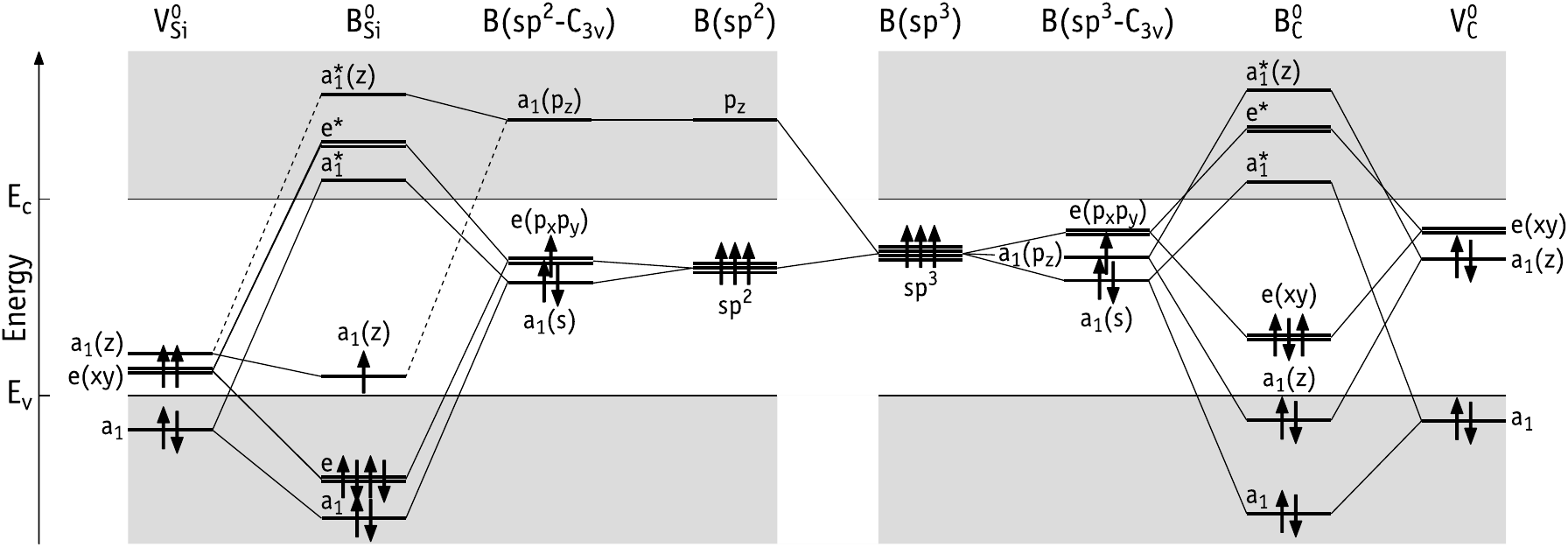}

\caption{\label{fig2}Schematic one-electron models of $\textrm{B}_{\textrm{Si}}^{0}$
(left) and $\textrm{B}_{\textrm{C}}^{0}$ (right) impurities in 4H-SiC
constructed by hybridization of valence states from sp$^{2}$ and
sp$^{3}$ atomic boron (middle-left and middle-right) with states
from the silicon and carbon vacancies (left and right), respectively.
Labeling of states is according to the $C_{3v}$ point group, except
for isolated atoms in the middle. Some labels include the direction
of the wave function polarization within parentheses. The diagrams
are spin-averaged with upward/downward arrows indicating the level
occupancy.}
\end{figure*}

The above threefold coordinated $\textrm{B}_{\textrm{Si}}$ defects
are markedly different from those found from previous local density
functional calculations. $\textrm{B}_{\textrm{Si}}$ in 3C-SiC was
essentially reported as a fourfold coordinated center, showing only
slightly different B-C bond lengths due to a weak JT driven $C_{3v}$
distortion \citep{Fukumoto1996,Bockstedte2001}. Four-fold coordination
was also found for $\textrm{B}_{\textrm{Si}}$ in 4H-SiC \citep{Deak2003}.
The neutral state was in this case interpreted as a shallow acceptor,
binding a diffuse hole with the character of an EMT state. These conclusions
are clearly at variance with our results --- we find that (1) the
paramagnetic $\textrm{B}_{\textrm{Si}}^{0}$ state is a singlet, showing
the highest symmetry allowed by the crystalline host, \emph{i.e.},
it is immune to the JT effect, and (2) is strongly localized on the
carbon radical next to boron, which is not in line with an EMT state.

An explanation for the above conflict was put forward by Gerstmann
\emph{et~al.} \citep{Gerstmann2004}, who interpreted the prediction
of an effective-mass character for $\textrm{B}_{\textrm{Si}}$ as
a failure of LDA, and as a corollary, a failure to describe the measured
$^{13}$C hyperfine data: “\emph{like the well-known underestimation
of the fundamental band gap, the localization of this defect state
is also strongly underestimated}”. Accordingly, the LDA gap is about
50\% narrower than the measured value, and for that reason, the C
dangling bond state becomes artificially over-mixed with the SiC valence
states. On the other hand, the non-local HSE06 functional predicts
a 3.2 eV wide-gap for 4H-SiC, allowing the singlet acceptor state
to emerge above the valence band top.

An analogous effect was found by Gouveia and Coutinho \citep{Gouveia2019}
for $\textrm{C}_{\textrm{i}}$ in 3C-SiC, but in this case involving
the mixing of a gap level with the conduction band. Based on analysis
of the Kohn-Sham data of structures ranging between the $D_{2d}$
(ground state with spin-1) and $C_{1h}$ (metastable and diamagnetic)
structures of $\textrm{C}_{\textrm{i}}^{0}$, an overestimated mixing
between the $\textrm{C}_{\textrm{i}}$ highest occupied level and
the conduction band states, was attributed to the narrow (semi\nobreakdash-)local
approximated band gap, which favored the incorrect $C_{1h}$ structure.
Besides the exchange-correlation treatment, this effect may depend
on other factors, most notably the dispersion of the defect state
(the mixing could be $\mathbf{k}$-point dependent), the sampling
of the BZ, or the size/shape of the supercells. The authors of Ref.~\citep{Gouveia2019}
used 512-atom cubic cells with the Brillowin zone sampled at $\Gamma$.
More recently, Schultz \emph{et~al.} \citep{Schultz2021} found that
upon improving the sampling to $2\times2\times2$ (using identical
supercells and GGA-level exchange-correlation treatment), the correct
spin-1 $D_{2d}$ state could be recovered. In Ref.~\citep{Oda2013}
it was noted that the $C_{1h}$ metastable structure of $\textrm{C}_{\textrm{i}}^{0}$,
was the most stable when using 216-atom cells with $\Gamma$-sampling
even at hybrid-DFT level. However, upon adding $\mathbf{k}$-points
away from $\Gamma$ to the sampling mesh (where the gap is wider),
the correct $D_{2d}$ configuration was also recovered. The result
of Ref.~\citep{Bockstedte2003}, where the $C_{1h}$ structure was
originally proposed as the most stable, was therefore attributed to
errors related to calculation settings.

One could ask if, for the case of B$_{\textrm{Si}}$, the above effect
results from poor sampling of the Brilloui zone. In Ref.~\citep{Supplemental}
we clearly demonstrate that the valence band edge overmixing of B$_{\textrm{Si}}$
at the GGA-level is a stable result and was found even for high-density
$\mathbf{k}$-point samplings (up to $4\times4\times4$).

From inspection of the band structure of defective supercells we arrived
at the orbital model for $\textrm{B}_{\textrm{Si}}^{0}$ depicted
on the left hand side of Fig.~\ref{fig2}. It consists of a schematic
diagram without spin resolution. Upward/downward arrows simply reflect
the electron occupancy. The model postulates how the $\textrm{sp}^{2\uparrow\uparrow\uparrow}$
states of atomic B(sp$^{2}$) unfold under the effect of a trigonal
crystal field of B(sp$^{2}$-$C_{3v}$), and how these hybridize with
the silicon vacancy states ($V_{\textrm{Si }}^{0}$) to produce the
electronic structure of $\textrm{B}_{\textrm{Si}}^{0}$. Accordingly,
three short B-C bonds of $\textrm{B}_{\textrm{Si}}$ are formed with
the participation of six electrons on low-energy bonding states $a_{1}+e$.
These result from overlap of $a_{1}$ and $e(xy)$ states localized
on three C atoms edging $V_{\textrm{Si }}^{0}$, with $a_{1}(\textrm{s})$
and $e(\textrm{p}_{x}\textrm{p}_{y})$ of threefold coordinated B(sp$^{2}$-$C_{3v}$).
Both $a_{1}+e$ and corresponding anti-bonding states $a_{1}^{*}+e^{*}$
of $\textrm{B}_{\textrm{Si}}^{0}$ are resonant with the valence and
conduction bands, respectively. The weak interaction between $a_{1}(z)$
(localized on the fourth carbon radical of $V_{\textrm{Si }}^{0}$)
with the $a_{1}(\textrm{p}_{z})$ state from the displaced boron atom,
leaves the former within the gap and semioccupied. The $a_{1}(z)$
state is the C radical responsible for the acceptor activity of $\textrm{B}_{\textrm{Si}}$,
the short covalent B-C bonds naturally explain the off-site distortion
without JT effect.

A picture close to that of Fig.~\ref{fig2} was discussed in the
literature nearly three decades ago by Bratus and co-workers \citep{Bratus1993}.
From analysis using a linear combination of atomic orbitals (LCAO),
it was argued that the $\textrm{sp}^{2}+\textrm{p}_{z}$ hybridization
of boron on the Si site was more stable than $\textrm{sp}^{3}$ simply
because (1) the covalent radius of B is much smaller than that of
Si and (2) the three bonds of B($\textrm{sp}^{2}$) with carbon ($\sim\!1.6$~Å
long) are considerably shorter than the host Si-C bonds ($\sim\!1.9$~Å).
Among the main conclusions was also the description of B$_{\textrm{Si}}^{0}$
ground state as a singlet, and consequently, that the observed displacement
of B from the perfect crystalline site could be explained without
a Jahn-Teller effect.

Subsequent studies of Petrenko \emph{et al.} \citep{Petrenko1996},
now using a semi-empirical modified neglect of diatomic overlap method,
also supported a pronounced off-site location for B$_{\textrm{Si}}$
in SiC. The crystalline host was approximated as a hydrogen saturated
spherical cluster of $\sim\!90$ SiC atoms (3C phase). With the emergence
of first-principles local density functional supercell calculations,
a fourfold coordinated structure for $\textrm{B}_{\textrm{Si}}$ with
effective-mass character became favored, suggesting that the findings
of Ref.~\citep{Petrenko1996} resulted from limitations of the method
employed. For instance, one could argue that due to quantum confinement
and underscreening effects, the band gap of the small clusters was
rather wide. That effect could have eliminated the mixing of $a_{1}(z)$
with the valence, thus favoring the $\textrm{sp}^{2}$-like bonding
of boron. Another argument cautioning against the off-site location
of B$_{\textrm{Si}}$ is the fact that such relaxations are often
overestimated when modeling defects in H-terminated clusters.

\begin{figure}
\includegraphics[clip,width=8.5cm]{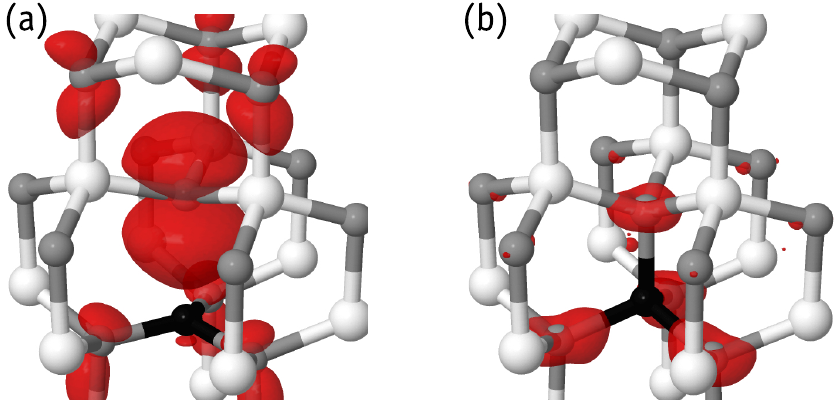}

\caption{\label{fig3}Spin density isosurface (cutoff 0.003~e/Å$^{3}$) of
neutral $\textrm{B}_{\textrm{Si}}$ defects at the $h$ site of 4H-SiC.
(a) Off-site threefold coordinated ground state configuration obtained
within HSE06. (b) Four-fold configuration as found from a GGA-level
calculation. Si, C and B atoms are shown in white, gray and black,
respectively.}
\end{figure}

On the contrary, we argue that the (semi-)local density functional
results for neutral B$_{\textrm{Si}}$ are spurious, that hybrid DFT
finds the correct off-site location of the B atom, and the LCAO-based
arguments of Bratus \emph{et~al.} \citep{Bratus1993} were essentially
correct after all. Figure~\ref{fig3} depicts the spin density in
the vicinity of neutral B$_{\textrm{Si}}(h)$ in 4H-SiC as found for
(a) the off-site ground state configuration within hybrid-DFT/HSE06
and (b) the on-site ground state configuration within conventional
DFT/GGA. Both isosurfaces have the same spin density cutoff (0.003~e/Å$^{3}$).
They depict the border within which the magnitude of the spin density
is above the specified threshold. Figure~\ref{fig3}(a) shows that
the amplitude of the spin density near the core of threefold coordinated
B$_{\textrm{Si}}^{0}$ is much larger than in the fourfold coordinated
configuration. In the latter case, many isosurface \emph{bubbles}
(with that specific spin density magnitude) are scattered across the
supercell volume, hidden behind the spheres and cylinders used to
represent atoms and bonds. Upon decreasing the cutoff by half, no
spin density isosurface could be seen for the fourfold coordinated
boron, while the p-like state of threefold coordinated boron was well
visible. This is consistent with deep threefold and shallow fourfold
states, respectively. Clearly, the DFT/GGA approximation predicts
a diffuse state with very little localization at the core of the defect.

\noindent 

Still regarding the bonding character of $\textrm{B}_{\textrm{Si}}^{0}$
in SiC, we note that this center is isovalent to substitutional nitrogen
on the Si site of SiC (N$_{\textrm{Si}}$) \citep{Deak1998} as well
as substitutional nitrogen in diamond (N$_{\textrm{s}}$) \citep{Smith1959}.
Within a simple Lewis picture, $\textrm{B}_{\textrm{Si}}^{0}$ can
be represented as $[\equiv\!\textrm{B}_{\textrm{Si}}\;\bullet\!\textrm{C}\!\equiv]$,
where each horizontal bar stands for a single C-B or C-Si bond, and
the bullet is an umparied electron. Analogously, N$_{\textrm{Si}}^{0}$
in SiC and neutral substitutional N in diamond can be described as
$[\equiv\!\textrm{N}_{\textrm{Si}}\!:\;\bullet\textrm{C}\!\equiv]$
and $[\equiv\!\textrm{N}_{\textrm{s}}\!:\;\bullet\textrm{C}\!\equiv]$,
respectively, where the dots “:” represent a lone-pair of electrons
tightly bound to nitrogen and deep within the valence band. Like the
B species in the Si site of SiC, N atoms with four carbon nearest
neighbors become threefold coordinated next to a paramagnetic C radical.
However, unlike B$_{\textrm{Si}}$, local and semilocal density functional
calculations account well for their off-site structure \citep{Deak1998,Bockstedte2004b,Jones2009}.
Although an explanation for such behavior is outside the scope of
the present work, we speculate that short C-N bonds combined with
Coulomb repulsion between the N lone pair and the unpaired electron
on the C dangling bond could be important ingredients for the stabilization
of the off-site configuration. A strong indication in favor of this
argument is that while the C-radical of $\textrm{B}_{\textrm{Si}}^{0}$
in SiC induces a semioccupied state low in the gap, C-radicals of
N$_{\textrm{Si}}$ in SiC and N$_{\textrm{s}}$ in diamond lead to
semioccupied states in the upper half of the gap, suggesting a stronger
repulsion of the unpaired electron in the N-related defects.

The semioccupied $a_{1}^{\uparrow}(z)$ singlet of $\textrm{B}_{\textrm{Si}}^{0}$
in 4H-SiC is represented in Fig.~\ref{fig2} just above the valence
band top. A spin-averaged calculation of $\textrm{B}_{\textrm{Si}}^{0}(h_{\textrm{a}})$
reveals that this level is located 0.52 eV above the highest occupied
Kohn-Sham level from the bulk. On the other hand, in a spin-polarized
calculation the spin-up $a_{1}^{\uparrow}(z)$ level lies within the
valence band (the highest occupied state is bulk-like), while the
spin-down component of $a_{1}(z)$ is 1.47~eV above the $E_{\textrm{v}}$
level. This picture is indicative of deep acceptor activity.

Upon atomic relaxation of negatively charged defects ($\textrm{B}_{\textrm{Si}}^{-}$),
we found that independently of the lattice site and initial configuration,
the boron atom moved to the perfect substitutional site, thus forming
four nearly equivalent 1.77~Å long B-C bonds. Concurrently, the $a_{1}(\textrm{p}_{z})$
state of boron increased its mixing with $a_{1}(z)$ from $V_{\textrm{Si}}$
to form the fourth B-C bond. The resulting $a_{1}(z)$ bond state
from $\textrm{B}_{\textrm{Si}}^{-}$ became resonant with the valence,
and the Kohn-Sham band gap was left clean. This does not imply that
B$_{\textrm{Si}}^{-}$ cannot capture a hole to become neutral. It
does not imply that it is a shallow acceptor either. As will be shown
in Sec.~\ref{subsec:dlts}, hole capture is accompanied by reconfiguration
to the threefold coordinated structure, making the hole trap relatively
deep. As summarized in Fig.~\ref{fig1}(c), the energy of $\textrm{B}_{\textrm{Si}}^{-}(h)$
was found slightly lower (0.04~eV) than that of $\textrm{B}_{\textrm{Si}}^{-}(k)$.

\begin{figure}
\includegraphics[clip,width=8.5cm]{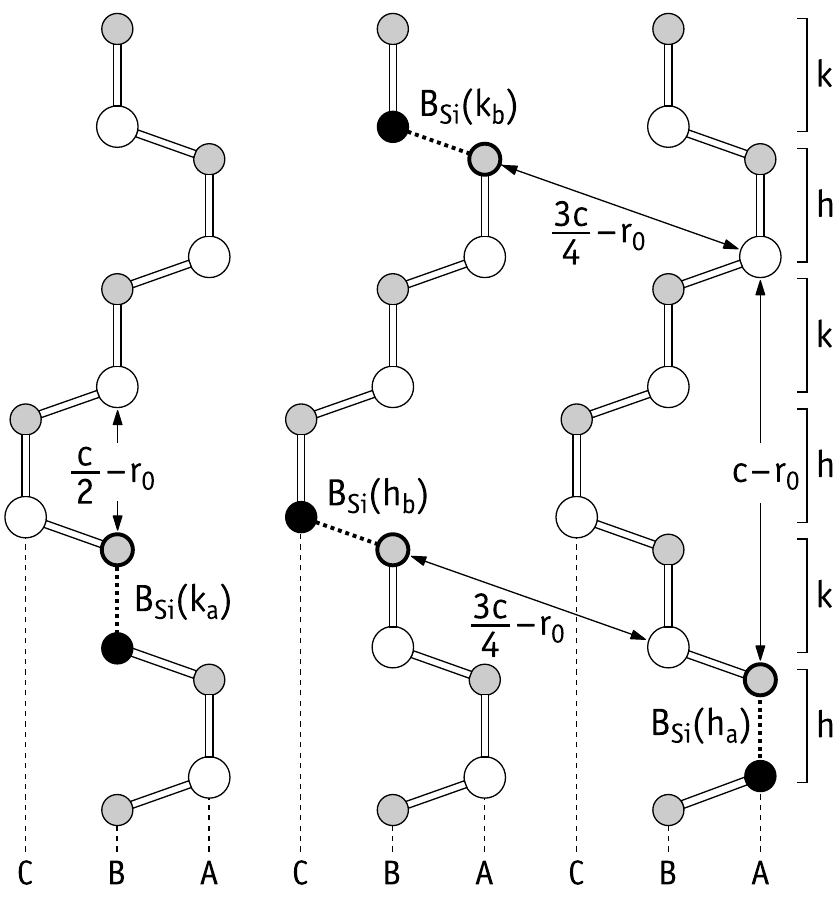}

\caption{\label{fig4}Location and possible alignments of $\textrm{B}_{\textrm{Si}}$
defects on the $\{11\bar{2}0\}$ plane of 4H-SiC. Silicon and carbon
atoms are shown in white and gray. Boron and carbon at the core of
the defect are represented as black and gray-haloed circles. For the
sake of clarity, all atomic positions are those of the perfect crystal.
The C-B broken bond of the neutral state is represented as a dotted
line. Relevant distances are indicated next to arrows, where $c$
and $r_{0}$ are the axial lattice parameter and the Si-C bond length,
respectively. Stacking (A, B, C) and site ($k$, $h$) indexes are
also indicated.}
\end{figure}

Now we look at the origin of the site-dependent alignment of $\textrm{B}_{\textrm{Si}}^{0}$
in 4H-SiC (the arguments discussed below apply to other polytypes
as well). The analysis is best followed with help of Fig.~\ref{fig4}.
In 4H-SiC, the stacking of SiC dimers along the $c$ axis occurs according
to a A-B-C-B sequence, where A and C are hexagonal bilayers and B
are cubic bilayers. Importantly, while hexagonal SiC dimers (type
A and C) are replicated in steps of length $c$ along the main crystallographic
direction (where $c$ is the axial lattice parameter), cubic bilayers
(type B) are repeated every $c/2$-long steps. This results in a \emph{wavier}
electrostatic potential and a stronger electric field in crystalline
regions along type B columns (see Fig.~\ref{fig4}).

The $a_{1}^{\uparrow}(z)$ state on the C radical of $\textrm{B}_{\textrm{Si}}^{0}(k_{\textrm{a}})$
interacts with the extensive $3\textrm{sp}^{3}$ valence electrons
of the nearest Si atom along the axis, only $c/2-r_{0}=3.14$~Å away
from carbon, where $r_{0}$ is the Si-C bond length (see left hand
side of Fig.~\ref{fig4}). This repulsion effectively raises the
energy of $\textrm{B}_{\textrm{Si}}^{0}(k_{\textrm{a}})$ by 0.11
eV with respect to $\textrm{B}_{\textrm{Si}}^{0}(h_{\textrm{a}})$.
In the latter case, the Si atom on the back of the $\textrm{C}\cdots\textrm{B}_{\textrm{Si}}(h_{\textrm{a}})$
unit is $c-r_{0}=8.17$~Å away from C (see right hand side of Fig.~\ref{fig4}).

\noindent 

Due to symmetry reasons, the above analysis cannot be strictly applied
to $\textrm{B}_{\textrm{Si}}^{0}(k_{\textrm{b}})$ and $\textrm{B}_{\textrm{Si}}^{0}(h_{\textrm{b}})$
defects (with C radicals polarized along Si-C basal bonds). However,
analogous conclusions may be drawn by inspecting the amount of empty
space between the carbon radical and the nearest atom along the $\textrm{B}\cdots\textrm{C}$
direction. As depicted in the middle of Fig.~\ref{fig4}, in a pristine
4H-SiC crystal, that distance is $3c/4-r_{0}=5.64$~Å for both $k$
and $h$ sites, thus lying right between the lower and upper limits
of the axially distorted configurations. This is consistent with the
energy ordering found for $\textrm{B}_{\textrm{Si}}^{0}(h_{\textrm{a}})<\textrm{B}_{\textrm{Si}}^{0}(k_{\textrm{b}})\sim\textrm{B}_{\textrm{Si}}^{0}(h_{\textrm{b}})<\textrm{B}_{\textrm{Si}}^{0}(k_{\textrm{a}})$.

The reorientation barrier between basal and axial distortions of neutral
$\textrm{B}_{\textrm{Si}}$ defects, was found from a batch of nudged
elastic band (NEB) calculations encompassing five intermediate structures
between initial and final states. See Ref.~\citep{Supplemental}
(and also Ref.~\citep{Henkelman2000}) for details of the barrier
calculations. From the results we find activation barriers of 0.04~eV
and 0.06~eV for $k_{\textrm{a}}\rightarrow k_{\textrm{b}}$ and $h_{\textrm{b}}\rightarrow h_{\textrm{a}}$
reorientations. These jumps involve a return from metastable to lowest
energy structures of $\textrm{B}_{\textrm{Si}}^{0}$ in $k$ and $h$
sites, respectively. These figures are reflected in the diagram of
Fig.~\ref{fig1}(c). Given the above meV-range barriers, the metastable
states are probably not formed, even at liquid-He temperature.

The reorientation of the C radical of $\textrm{B}_{\textrm{Si}}(k)$
between equivalent basal orientations was also investigated using
the NEB method. We found that $\textrm{B}_{\textrm{Si}}(k)$ has to
surmount a barrier of 0.09~eV to perform a $k_{\textrm{b}}\rightarrow k_{\textrm{b}'}$
jump between neighboring alignments with the same energy. Hence, above
a certain (low) temperature, $\textrm{B}_{\textrm{Si}}(k)$ is likely
to roam around all equivalent $k_{\textrm{b}}$ distortions, showing
effective thermally averaged $C_{3v}$ symmetry.

\subsection{Boron on the carbon site: deep boron\label{subsec:db}}

Regarding the boron replacement of carbon ($\textrm{B}_{\textrm{C}}$),
we found that the boron impurity sits very close to the crystalline
site. Very small B-C bond distortions were obtained when symmetry
breaking was allowed during the relaxations. From inspection of the
Kohn-Sham band structure we found that the on-site configuration (with
$C_{3v}$ symmetry) introduces a deep doublet state in the gap. In
a spin-averaged calculation of a trigonal $\textrm{B}_{\textrm{C}}^{0}(h)$
defect, a pair of doubly degenerate Kohn-Sham states occupied by three
electrons appear at 0.29~eV above the highest occupied level from
the bulk. On the other hand, in a spin-polarized calculation of the
same structure the spin-up $e^{\uparrow\uparrow}(xy)$ level lies
at 0.06~eV above $E_{\textrm{v}}$, whereas the spin-down counterpart
$e^{\downarrow}(xy)$ is 0.44~eV above the $E_{\textrm{v}}$. Note
that these figures neglect any Jahn-Teller relaxation and electron-phonon
coupling effects (the occupation of the doublets was fixed --- not
variational).

A simplified bond orbital model for neutral $\textrm{B}_{\textrm{C}}$
is shown on the right half of Fig.~\ref{fig2}. It represents the
conversion of atomic boron B(sp$^{3}$) states under the effect of
a trigonal crystal field, B($\textrm{sp}^{3}$-$C_{3v}$), and the
hybridization of the later with $a_{1}^{\uparrow\downarrow}+a_{1}^{\uparrow\downarrow}(z)+e(xy)$
states of the carbon vacancy (where boron is sitting). The Si radicals
edging the $V_{\textrm{C}}$ defect are considerably more diffuse
than the C radicals in $V_{\textrm{Si}}$, and therefore their overlap
with boron is significant for all states. The result is the formation
of bonding $a_{1}^{\uparrow\downarrow}+a_{1}^{\uparrow\downarrow}(z)$
and anti-bonding $a_{1}^{*}+a_{1}^{*}(z)$ singlets within the valence
and conduction bands, respectively, while a partially occupied $e^{\uparrow\downarrow\uparrow}(xy)$
doublet is left in the gap. The $a_{1}$ and $a_{1}(z)$ states are
respectively located on basal and axial B-Si bonds, while the components
of $e(xy)$ are B-centered $\textrm{p}_{x}$- and $\textrm{p}_{y}$-like
states overlapping basal bonds only. It is clear that any electronic
activity of $\textrm{B}_{\textrm{C}}$ must be ascribed to the $e(xy)$
state.

Upon monoclinic distortion ($C_{1h}$ symmetry), the $e^{\uparrow\downarrow\uparrow}(xy)$
neutral state can either split into $a''^{\uparrow\downarrow}(x)+a'{}^{\uparrow}(y)$
or $a'^{\uparrow\downarrow}(y)+a''{}^{\uparrow}(x)$ states with net
spin $S=1/2$. Here $a'$ and $a''$ are respectively symmetric and
anti-symmetric with respect to a $\{2\bar{1}\bar{1}0\}$ mirror plane.
While $a''$ is a $\textrm{p}_{x}$-like state with a node coincident
with the mirror plane, $a'$ is $\textrm{p}_{y}$-like with a node
on the boron atom and polarized along $\langle01\bar{1}0\rangle$.
Irrespectively of the lattice site, we found that the most stable
JT-distorted configuration of B$_{\textrm{C}}^{0}$ involved a minute
($\sim\!0.06$~Å) displacement of boron along $\langle01\bar{1}0\rangle$,
leading to two shorter B-Si bonds (and a slightly elongated one).
That configuration corresponds to the electronic state $a'^{\uparrow\downarrow}+a''{}^{\uparrow}$.
The alternative $a''^{\uparrow\downarrow}+a'{}^{\uparrow}$ state
was metastable by 15~meV only. In overall, $\textrm{B}_{\textrm{C}}^{0}(k)$
was more stable than $\textrm{B}_{\textrm{C}}^{0}(h)$ by 39 meV.

Interestingly, and despite the minute JT-driven bond deformations,
the relaxation energy with respect to the high-symmetry ($C_{3v}$)
state was about 0.25~eV for both $\textrm{B}_{\textrm{C}}^{0}(k)$
and $\textrm{B}_{\textrm{C}}^{0}(h)$. This is a surprisingly large
value, and as far as we could find, it is not an artifact. The electronic
occupancy of the high symmetry state (at the JT singularity) was not
variational during the self-consistent cycle, and each pair of spin
components of the doublet kept equal occupancy.

While the JT relaxation energy is a considerable barrier to surmount
at liquid-He temperature, the question is --- how likely is boron
able to jump between neighboring off-axis configurations and show
a dynamic Jahn-Teller effect? There are in total 6 possible JT displacements
of boron away from the perfect C-site. They comprise alternating $a'^{\uparrow\downarrow}+a''{}^{\uparrow}$
and $a''^{\uparrow\downarrow}+a'{}^{\uparrow}$ states around the
hexagonal axis of the crystal, dephased by a rotation angle of $\pi/3$.
Jumping between neighboring structures involves a displacement of
the B atom of only 0.04~Å. Although the barrier was not calculated
with a proper transition-state method, it was estimated from the energy
of the structure at mid-way between two neighboring $\textrm{B}_{\textrm{C}}^{0}(k)$
and $\textrm{B}_{\textrm{C}}^{0}(h)$ states. The small traveling
distance of the B atom justifies this simple approach. Accordingly,
we found that the rotation barrier is about 15~meV for both $\textrm{B}_{\textrm{C}}^{0}(k)$
and $\textrm{B}_{\textrm{C}}^{0}(h)$. Such minute figure is smaller
than the zero-point energy of an oscillating B-Si bond, suggesting
that the $\textrm{B}_{\textrm{C}}^{0}$ defects effectively roam around
the $c$ axis, thus showing a dynamic-JT effect even at liquid-helium
temperature.

In the negative charge state, the doublet becomes fully occupied and
$\textrm{B}_{\textrm{C}}^{-}$ recovers the full trigonal symmetry
of the C-site. In this charge state, the impurity at the $k$-site
is 76~meV more stable than at the $h$-site.

\subsection{Connection with optical and junction spectroscopy\label{subsec:dlts}}

\begin{figure}
\includegraphics[clip,width=8.5cm]{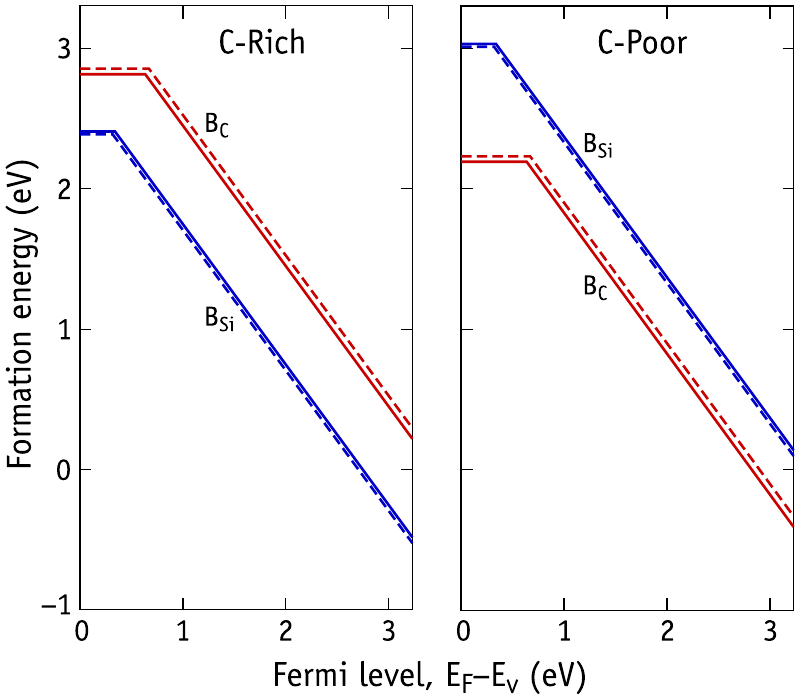}

\caption{\label{fig5}Formation energy diagrams of $\textrm{B}_{\textrm{Si}}$
(blue lines) and $\textrm{B}_{\textrm{C}}$ (red lines) in 4H-SiC.
Solid and dashed lines represent formation energies of boron defects
located at $k$ and $h$ sites, respectively.}
\end{figure}

The formation energy of boron impurities, obtained according to Eq.~\ref{eq:eform},
is shown in Fig.~\ref{fig5}. There we show the results for the formation
energy of $\textrm{B}_{\textrm{Si}}$ and $\textrm{B}_{\textrm{C}}$
defects in 4H-SiC under carbon rich and poor conditions (left- and
right-hand side diagrams, respectively), as a function of the Fermi
energy (referred with respect to the valence band top). Solid and
dashed lines refer to boron defects located at $k$ and $h$ sites,
respectively.

\noindent 

Clearly, and in agreement with previous findings \citep{Fukumoto1996,Bockstedte2001},
in carbon rich material, where depletion of Si is favored, $\textrm{B}_{\textrm{Si}}$
has lower formation energy than $\textrm{B}_{\textrm{C}}$. The opposite
is found for C-poor material. At growth temperatures, where the Fermi
level can be assumed to be at mid-gap, the formation energy of $\textrm{B}_{\textrm{Si}}$
is 0.74-0.85~eV lower than that of $\textrm{B}_{\textrm{C}}$ in
C-rich samples. On the other hand, $\textrm{B}_{\textrm{C}}$ is more
stable than $\textrm{B}_{\textrm{Si}}$ by 0.36-0.48~eV in C-poor
samples. The ranges result from considering $k$ and $h$ sites for
each impurity.

Figure~\ref{fig5} shows that both $\textrm{B}_{\textrm{Si}}$ and
$\textrm{B}_{\textrm{C}}$ are single acceptors. The defects adopt
a negative charge state for a wide range of Fermi levels, and we did
not find donor transitions or additional acceptor transitions within
the gap.

Considering the lowest-energy configurations of neutral $\textrm{B}_{\textrm{Si}}$
defects at $k$ and $h$ sites, we place the acceptor levels of $\textrm{B}_{\textrm{Si}}(k)$
and $\textrm{B}_{\textrm{Si}}(h)$ at $E_{\textrm{v}}+0.34$~eV and
$E_{\textrm{v}}+0.32$~eV, respectively. These results are shown
graphically in Fig.~\ref{fig1}(c), and they indicate that the binding
energy of the hole to $\textrm{B}_{\textrm{Si}}$ is almost independent
of the lattice site, despite the adoption of rather distinct crystalline
alignments by neutral $\textrm{B}_{\textrm{Si}}(k_{\textrm{b}})$
and $\textrm{B}_{\textrm{Si}}(h_{\textrm{a}})$ ground states.

These results are in line with the observation of a single peak by
DLTS and Laplace-DLTS related to a hole trap of shallow boron at $E_{\textrm{v}}+0.27$~eV
\citep{Suttrop1990,Storasta2002,Capan2020}. Despite the agreement,
we note that the calculated difference between the acceptor levels
of $\textrm{B}_{\textrm{Si}}$ at $k$ and $h$ site (20~meV), is
smaller than the typical error of the method employed for the calculation.
Additionally, the detection of a single peak by the Laplace-DLTS technique
suggests that the difference could be even smaller, or that one of
the configurations is dominant. The calculated relative energies of
$\textrm{B}_{\textrm{Si}}(k)$ and $\textrm{B}_{\textrm{Si}}(h)$
do not support the second possibility.

\noindent 
\begin{figure}
\includegraphics[clip,width=8.5cm]{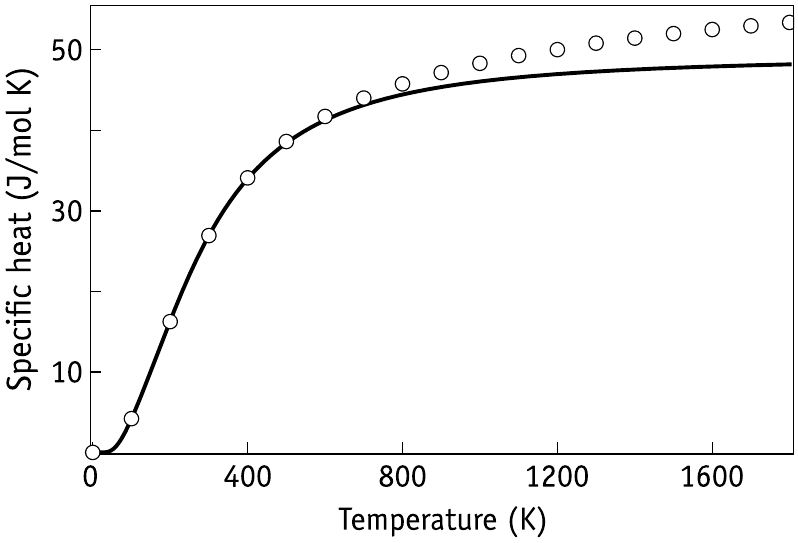}

\caption{\label{fig6}Specific heat of 4H-SiC calculated at constant volume
within the harmonic approximation (solid line). Circles represent
measured data for $\alpha$-SiC, obtained under constant pressure
conditions as reported in Ref.~\citep{Chase1998}. The calculation
employed Eq.~\ref{eq:cv} and considered a total of 213 vibrational
frequencies from a 72-atom 4H-SiC supercell.}
\end{figure}

An important question relates to the mechanism behind the capture
of holes by $\textrm{B}_{\textrm{Si}}^{-}$. After all, the band structure
of a supercell with this defect state shows a clean band gap. Our
findings indicate that the mechanism involves a strong electron-phonon
coupling, much like in a polaronic trapping effect \citep{Stoneham2007}.
Essentially, the off-site distortion of $\textrm{B}_{\textrm{Si}}^{-}$
raises an occupied level above the valence band top, which is then
stabilized upon hole capture. The first stage (level raising above
$E_{\textrm{v}}$) translates into the surmounting of a capture barrier,
estimated to be of the order of 0.1~eV. See Ref.~\citep{Supplemental}
(and also Refs.~\citep{Stoneham1981,Alkauskas2008,Wang2009,Alkauskas2012,Alkauskas2016})
for details regarding the raising of the level above $E_{\textrm{v}}$
and the estimation of the capture barrier.

Regarding boron on the carbon site, we find $(-/0)$ transitions at
$E_{\textrm{v}}+0.63$~eV and $E_{\textrm{v}}+0.67$~eV for $\textrm{B}_{\textrm{C}}(k)$
and $\textrm{B}_{\textrm{C}}(h)$, respectively. Neutral ground states
with electronic configuration ${a'}^{\uparrow\downarrow}+{a''}^{\uparrow}$
were considered in our calculations. These figures agree well with
early and recent measurements in 6H- and 4H-SiC \citep{Anikin1985,Suttrop1990,Sridhara1998,Storasta2002,Capan2020},
which indicate a transition of deep boron in the range 0.5-0.7~eV
above the valence band top.

\noindent 
\begin{figure}
\includegraphics[clip,width=8.5cm]{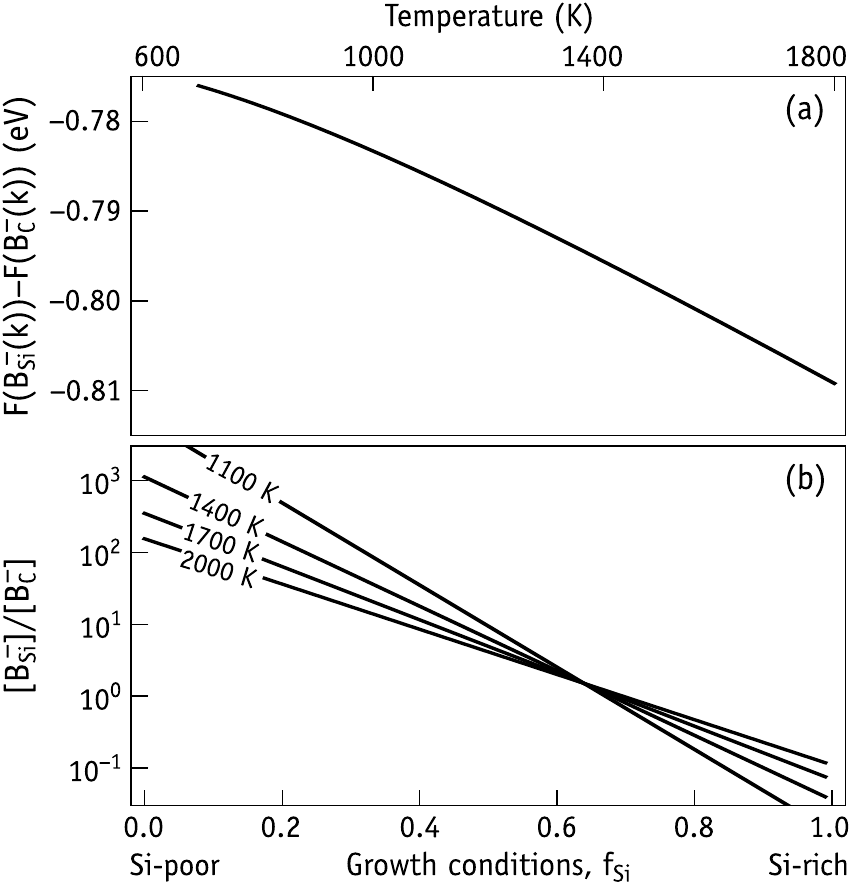}

\caption{\label{fig7}(a) Free energy of $\textrm{B}_{\textrm{Si}}^{-}(k)$
with respect to that of $\textrm{B}_{\textrm{C}}^{-}(k)$ in 4H-SiC
as a function of temperature under Si-poor conditions. The Fermi level
was considered to be located at midgap. (b) Concentration ratio of
$\textrm{B}_{\textrm{Si}}^{-}$ to that of $\textrm{B}_{\textrm{C}}^{-}$
as a function of the stoichiometric growth conditions (represented
by $f_{\textrm{Si}}$), for selected temperatures.}
\end{figure}

The separation between calculated levels of $\textrm{B}_{\textrm{C}}(k)$
and $\textrm{B}_{\textrm{C}}(h)$ is small, $\approx40$~meV, but
about twice larger than the analogous figure obtained for $\textrm{B}_{\textrm{Si}}$.
Again, this difference is lower than the error of the calculations,
and therefore should be considered with due care. Considering that
the signal of the D center was recently shown to comprise two equally
intense peaks separated by nearly 0.1~eV, our results support the
view that these peaks arise from two nearly equivalent deep boron
acceptors: a “shallower” configuration sitting at the cubic carbon
site and a “deeper” one replacing the hexagonal site. These correspond
to measured transitions at $E_{\textrm{v}}+0.49$~eV and $E_{\textrm{v}}+0.57$~eV,
respectively \citep{Capan2020}.

\subsection{Finite temperature calculations\label{subsec:ft}}

Up until now, our results refer to zero temperature conditions, not
even accounting for differences in zero-point motion between $\textrm{B}_{\textrm{Si}}$
and $\textrm{B}_{\textrm{C}}$ species. However, at high temperatures
the effect of entropy to the relative stability of $\textrm{B}_{\textrm{Si}}$
and $\textrm{B}_{\textrm{C}}$ can be relevant. To strengthen our
conclusions, we evaluated their respective free energies of formation
at high temperatures, in particular under intrinsic conditions. For
the sake of testing the methodology we calculated the specific heat
at constant volume for bulk 4H-SiC as,

\begin{equation}
c_{\textrm{v}}(T)=-T\left(\frac{\partial^{2}F_{\textrm{vib}}}{\partial T^{2}}\right),\label{eq:cv}
\end{equation}
and the result is shown in Fig.~\ref{fig6}. In that plot, we also
report several data points recorded during experiments at constant
pressure for $\alpha$-SiC (6H-SiC) \citep{Chase1998}.

The calculated specific heat describes the measurements very well
up to nearly $T\!\sim\!800$~K, when anharmonic effects start to
gain importance, and beyond which the calculated free energy and its
derivatives become more qualitative. In Ref.~\citep{Gomes2022},
we demonstrated that these calculations cannot be improved by enlarging
the supercells. Also important, is the fact that the constant volume
calculations match well the constant pressure measurements across
a wide range of temperatures. The reason is hinted by the minute thermal
expansion of crystalline SiC, which is about $5\times10^{-6}\:\textrm{K}^{-1}$
for temperatures as high as 1000~$^{\circ}$C \citep{Stockmeier2009}.

The calculated difference in the free energy of formation $\Delta F(k)=F(\textrm{B}_{\textrm{Si}}^{-}(k))-F(\textrm{B}_{\textrm{C}}^{-}(k))$,
is shown in Fig.~\ref{fig7}(a) in the temperature range $T=600\textrm{-}1800$~K.
The quantity represented refers to impurities located in cubic sites.
For boron defects at the hexagonal sites the $T$-dependence of the
analogous quantity was almost identical, although its magnitude increased
by about 0.1~eV. Figure~\ref{fig7}(a) shows that $\textrm{B}_{\textrm{Si}}^{-}$
increases its relative stability with respect to $\textrm{B}_{\textrm{C}}^{-}$
by almost 0.05~eV when raising the temperature from 1000~K to 2000~K.
The implication of this result is illustrated in Fig.~\ref{fig7}(b)
where we plot the concentration ratio of $\textrm{B}_{\textrm{Si}}^{-}$
to $\textrm{B}_{\textrm{C}}^{-}$ defects as a function of the stoichiometric
conditions (represented by $f_{\textrm{Si}}$), at different temperatures.
Under equilibrium, the concentration ratio is given by

\begin{equation}
\frac{[\textrm{B}_{\textrm{Si}}^{-}]}{[\textrm{B}_{\textrm{C}}^{-}]}=\exp\left(-\frac{\Delta F(k)-\Delta F(h)}{k_{\textrm{B}}T}\right),\label{eq:ratio}
\end{equation}
where $\Delta F(k)$ and $\Delta F(h)$ are free energy differences
$F(\textrm{B}_{\textrm{Si}}^{-})-F(\textrm{B}_{\textrm{C}}^{-})$
pertaining $k$ and $h$ sites, respectively {[}as represented in
Fig.~\ref{fig7}(a){]}. For chemical vapor deposition grown material,
reactors typically run at temperatures of about 1600-1650~$^{\circ}$C
($T\!\sim\!1900$~K) \citep{Ito2008}. Under these conditions we
estimate $[\textrm{B}_{\textrm{Si}}^{-}]/[\textrm{B}_{\textrm{C}}^{-}]\approx200$
and about $0.1$ for a Si-poor and Si-rich stoichiometry, respectively.
It is evident that even for the limit of Si-poor growth, which is
the most favorable for introduction of the $\textrm{B}_{\textrm{Si}}$,
thermodynamics imposes the formation of deep boron centers with a
concentration about two orders of magnitude below that of the shallow
counterpart. Figure~\ref{fig7}(b) shows that even at $T=1700$~K
the $[\textrm{B}_{\textrm{Si}}^{-}]/[\textrm{B}_{\textrm{C}}^{-}]$
ratio is nearly 350, and probably the elimination of $\textrm{B}_{\textrm{C}}$
cannot be achieved during growth.

\noindent 
\begin{table}
\begin{ruledtabular}
\caption{\label{tab1}Calculated principal values ($g_{1}$, $g_{2}$ and $g_{3}$)
of the gyromagnetic $g$ tensor of paramagnetic boron-related defects
at $k$ and $h$ sites of 4H-SiC. Assignments of experimental values
from EPR signals of shallow (top) and deep (bottom) boron centers
are indicated by their location in the table. For trigonal states
($C_{3v}$), $g_{3}$ is parallel to the main crystallographic $c$
axis. For monoclinic states ($C_{1h}$), $g_{1}$ is perpendicular
to the $\{2\bar{1}\bar{1}0\}$ symmetry plane, while $g_{2}$ and
$g_{3}$ are rotated by an angle $\theta$ away from $\langle0\bar{1}10\rangle$
and $\langle0001\rangle$ directions, respectively. Dynamic trigonal
states (labeled with a subscripted “dyn”) refer to averaged $g$ tensors
involving three symmetrically equivalent $C_{1h}$ states (see text).
Also indicated are the temperatures of the measurements.}

\noindent %
\begin{tabular}{lcccccc}
 & $T$ (K) & Sym & $g_{1}$ & $g_{2}$ & $g_{3}$ & $\theta$ ($^{\circ}$)\tabularnewline
\hline 
$\textrm{B}_{\textrm{Si}}^{0}(k_{\textrm{b}})$ &  & $C_{1h}$ & 2.0068 & 2.0078 & 2.0028 & 70\tabularnewline
EPR \citep{GreulichWeber1997} & 4.2-45 & $C_{1h}$ & 2.0059 & 2.0069 & 2.0025 & 69\tabularnewline[0.4cm]
$\textrm{B}_{\textrm{Si}}^{0}(k_{\textrm{dyn}})$ &  & $C_{3v}$ & 2.0051 & 2.0051 & 2.0073 & 0\tabularnewline
EPR \citep{GreulichWeber1997} & 61-83 & $C_{3v}$ & 2.0046 & 2.0046 & 2.0064 & 0\tabularnewline[0.4cm]
$\textrm{B}_{\textrm{Si}}^{0}(h_{\textrm{a}})$ &  & $C_{3v}$ & 2.0089 & 2.0089 & 2.0022 & 0\tabularnewline
EPR \citep{GreulichWeber1997} & 4.2-83 & $C_{3v}$ & 2.0070 & 2.0070 & 2.0019 & 0\tabularnewline
\hline 
$\textrm{B}_{\textrm{Si}}(k)\textrm{-}V_{\textrm{C}}^{0}(k)$ &  & $C_{1h}$ & 2.0041 & 2.0065 & 2.0028 & 78\tabularnewline
$\textrm{B}_{\textrm{Si}}(k)\textrm{-}V_{\textrm{C}}^{0}(k_{\textrm{dyn}})$ &  & $C_{3v}$ & 2.0035 & 2.0035 & 2.0063 & 0\tabularnewline[0.4cm]
$\textrm{B}_{\textrm{C}}^{0}(k)$ &  & $C_{1h}$ & 2.0050 & 2.0205 & 2.0279 & 13\tabularnewline
$\textrm{B}_{\textrm{C}}^{0}(k_{\textrm{dyn}})$ &  & $C_{3v}$ & 2.0129 & 2.0129 & 2.0275 & 0\tabularnewline
EPR \citep{Baranov1998} & 4 & $\sim C_{3v}$ & 2.0 & 2.0 & 2.029 & $\sim0$\tabularnewline[0.4cm]
$\textrm{B}_{\textrm{C}}^{0}(h)$ &  & $C_{1h}$ & 2.0056 & 2.0173 & 2.0246 & 13\tabularnewline
$\textrm{B}_{\textrm{C}}^{0}(h_{\textrm{dyn}})$ &  & $C_{3v}$ & 2.0116 & 2.0116 & 2.0240 & 0\tabularnewline
EPR \citep{Baranov1998} & 4 & $\sim C_{3v}$ & 2.0 & 2.0 & 2.024 & $\sim0$\tabularnewline
\end{tabular}
\end{ruledtabular}

\end{table}

We finally note that from the calculated vibrational mode frequencies,
we could not find boron-related modes outside the spectrum of the
crystalline density of states. Therefore any boron vibrational mode
must be resonant, and most certainly hard to detect experimentally.

\subsection{Connection with EPR\label{subsec:epr}}

Figure~\ref{fig1} readily explains the rather distinct EPR signals
of shallow boron at $k$ and $h$ sites, as well as their temperature
dependence \citep{Zubatov1985}. While $\textrm{B}_{\textrm{Si}}^{0}(h)$
finds its ground state forming a paramagnetic p-like orbital on the
C atom of a broken B-C bond along the main crystalline axis, the $\textrm{B}_{\textrm{Si}}^{0}(k)$
lowest energy configuration has an analogous p-orbital (and a B-C
broken bond) but it is now along the direction of a basal bond of
the crystal.

The upper part of Tab.~\ref{tab1} records the calculated $g$ tensors
of shallow $\textrm{B}_{\textrm{Si}}^{0}$ defects in 4H-SiC, along
with the corresponding quantities measured by EPR \citep{GreulichWeber1997}.
For trigonal states ($C_{3v}$ symmetry), the main $g_{3}$ component
is assumed to be parallel to the main crystallographic $c$ axis.
For monoclinic states ($C_{1h}$ symmetry), $g_{1}$ is perpendicular
to the $\{2\bar{1}\bar{1}0\}$ symmetry plane, while $g_{2}$ and
$g_{3}$ are rotated by an angle $\theta$ away from $\langle0\bar{1}10\rangle$
and $\langle0001\rangle$ directions, respectively. Figures~\ref{fig1}(a)
and \ref{fig1}(b) show this convention graphically for $\textrm{B}_{\textrm{Si}}^{0}(k_{\textrm{b}})$
with a broken B-C bond on the $(2\bar{1}\bar{1}0)$ mirror plane and
for $\textrm{B}_{\textrm{Si}}^{0}(h_{\textrm{a}})$, respectively.

Ground-states $\textrm{B}_{\textrm{Si}}^{0}(k_{\textrm{b}})$ and
$\textrm{B}_{\textrm{Si}}^{0}(h_{\textrm{a}})$ have a calculated
main $g$ tensor component $g_{3}\!\sim\!2.002$ along the C radical,
making an angle with the $[0001]$ direction of $\theta=70{}^{\circ}$
and $0^{\circ}$, respectively (see also Fig.~\ref{fig1}). The $g$
tensors are nearly or perfectly axial for both static $\textrm{B}_{\textrm{Si}}^{0}(k_{\textrm{b}})$
and $\textrm{B}_{\textrm{Si}}^{0}(h_{\textrm{a}})$ structures, resulting
from the conspicuous alignment of the spin density on the carbon radical
as Fig.~\ref{fig3}(a) clearly displays. The match with the measurements
carried out at low temperature ($T\apprle45$~K) is excellent, both
in terms of magnitude (error $\apprle0.0005$) and monoclinic angle
(error $\sim\!1^{\circ}$). The error bar of the components perpendicular
to the C-radical ($g_{1}$ and $g_{2}$) is about 4 times larger,
but still, the agreement is deemed very good, especially considering
that both calculated and observed $g$ values show identical trends
in terms of axial character and anisotropy: $g_{2}-g_{1}=0.0010$
for $\textrm{B}_{\textrm{Si}}^{0}(k_{\textrm{b}})$, and $g_{3}-g_{1}=-0.0067$
for $\textrm{B}_{\textrm{Si}}^{0}(h_{\textrm{a}})$.

The calculated $g$ tensor components of Tab.~\ref{tab1} were found
by sampling the band structure over a $2\times2\times2$ mesh of $\mathbf{k}$-points.
A denser $3\times3\times3$-mesh calculation for $\textrm{B}_{\textrm{Si}}^{0}(h_{\textrm{a}})$
gave $g_{1}=g_{2}=2.0083$ and $g_{3}=2.0015$, which deviate from
the results with the coarser mesh by 0.0007. Most importantly, the
relative magnitude of the axial and transverse components is similar
in both calculations and match very well the observations.

As discussed at the end of Sec.~\ref{subsec:sb}, the activation
energy for rotation of the broken bond of $\textrm{B}_{\textrm{Si}}^{0}(k_{\textrm{b}})$
around $[0001]$ was estimated at about 0.1~eV, allowing the structure
to jump between all three equivalent alignments at rather low temperatures.
This result is consistent with the observed raise of symmetry of the
EPR signal assigned to shallow boron in cubic sites, from monoclinic
to trigonal above $T\sim\!45$~K. We argue that above this temperature,
the $\textrm{B}_{\textrm{Si}}^{0}(k)$ defect jumps between three
equivalent monoclinic configurations at a rate much faster than the
inverse of the EPR recording time. The result is the observation of
a “dynamic” state with effective $C_{3v}$ symmetry (hereafter labeled
with a “dyn” subscript), whose $g$ tensor is estimated by averaging
over all three equivalent $C_{1h}$ orientations. The calculated axial
component $g_{3}=2.0073$ of $\textrm{B}_{\textrm{Si}}^{0}(k_{\textrm{dyn}})$,
now along $[0001]$, mostly inherits contributions from $g_{2}=2.0078$
of static $\textrm{B}_{\textrm{Si}}^{0}(k_{\textrm{b}})$ configurations
{[}see Fig.~\ref{fig1}(a){]}, thus becoming the largest component.
This contrasts with $g_{3}$ of $\textrm{B}_{\textrm{Si}}^{0}(h_{\textrm{a}})$
which is the smallest component of this configuration. The magnitude
of the calculated $g$ values of $\textrm{B}_{\textrm{Si}}^{0}(k_{\textrm{dyn}})$
agrees very well with those assigned to shallow boron on the cubic
site measured in the temperature range $T=61\textrm{-}83$~K (error
$<0.001$). The calculated anisotropy $g_{3}-g_{1}\approx0.002$ for
$\textrm{B}_{\textrm{Si}}^{0}(k_{\textrm{dyn}})$ differs from the
measurements by $0.0004$ only.

The coupling of the unpaired spin of $\textrm{B}_{\textrm{Si}}^{0}$
defects with $^{13}$C and $^{11}$B magnetic isotopes quantifies
the magnitude and shape of the spin density at the core of the defect.
$^{11}$B and $^{13}$C hyperfine data was recorded experimentally
at 3.4~K \citep{Zubatov1985} and 1.5~K \citep{Matsumoto1997} by
EPR and ENDOR, respectively. Under these conditions $\textrm{B}_{\textrm{Si}}^{0}$
defects are static and the HF signals could be resolved. The calculated
principal values of the HF tensors due to interactions with $^{13}$C
and $^{11}$B elements at the broken C-B bond of $\textrm{B}_{\textrm{Si}}^{0}$
defects ($k_{\textrm{b}}$ and $h_{\textrm{a}}$ structures) are reported
in Tab.~\ref{tab2}. Also reported are the isotropic and anisotropic
HF constants ($a$ and $b$, respectively), which assume an axial
character for the wave function of the unpaired electron. The upper
and lower halves of the table show the results for boron located on
cubic and hexagonal sublattice sites, respectively. The experimental
data accompanying the calculations relate to boron defects in 6H-SiC
samples \citep{Zubatov1985,Matsumoto1997}.

The calculations confirm that the paramagnetic state is essentially
axial along direction 3 (see principal directions and monoclinic angle
$\theta$ in Fig.~\ref{fig1}). Differences between $A_{1}$ and
$A_{2}$ were always lower than 1~MHz. Both theory and experiments
indicate a relatively large and close $^{13}$C Fermi contact ($a\sim80\textrm{-}90$~MHz),
reflecting the large localization on the C radical. The calculated
anisotropic $^{13}$C HF constants ($b\sim50$~MHz) are also in fair
agreement with the EPR data ($b\sim40$~MHz). Although not statistically
meaningful, the error bar of the calculated HF constants (considering
the measurements reported in Tab.~\ref{tab2}) is estimated as $\lesssim10$~MHz.
We also note that the isotropic HF constants slightly underestimate
previous calculations based on the local density approximation (LDA)
\citep{Gerstmann2004}. This is interpreted as a tendency of GGA to
underlocalize the electron density in comparison to the overlocalization
of the LDA.

\noindent 
\begin{table}
\begin{ruledtabular}
\caption{\label{tab2}Calculated principal values of the hyperfine tensors
($A_{1}$, $A_{2}$ and $A_{3}$) for $^{13}$C and $^{11}$B species
located in the broken C-B bond of B$_{\textrm{Si}}^{0}(k)$ and B$_{\textrm{Si}}^{0}(h)$
defects in 4H-SiC. Isotropic and anisotropic HF constants ($a$ and
$b$) are also shown and they assume an axial state pointing along
direction 3. For the trigonal B$_{\textrm{Si}}^{0}(h)$ defect, directions
1 and 2 are along the basal plane. For the monoclinic B$_{\textrm{Si}}^{0}(k)$
defect, $B_{1}$ is the component perpendicular to the $\{2\bar{1}\bar{1}0\}$
mirror plane, while $B_{2}$ and $B_{3}$ are rotated by an angle
$\theta$ away from $\langle0\bar{1}10\rangle$ and $\langle0001\rangle$
directions, respectively (see Fig.~\ref{fig1}). EPR data from $^{13}$C-enriched
samples ($^{13}$C-EPR) \citep{Zubatov1985} and $^{11}$B-ENDOR \citep{Matsumoto1997}
are also included for comparison. These were obtained in 6H-SiC and
pertain to boron defects at $k_{1}$ ($k_{1}$ and $k_{2}$ data are
very similar) and $h$ sites. All HF couplings are in MHz.}

\noindent %
\begin{tabular}{lcccccc}
Defect & $A_{1}$ & $A_{2}$ & $A_{3}$ & $a$ & $b$ & $\theta$ ($^{\circ}$)\tabularnewline
\hline 
$^{13}$C-$\textrm{B}_{\textrm{Si}}(k_{\textrm{b}})$ & 34 & 34 & 183 & 84 & 50 & 72\tabularnewline
$^{13}$C-EPR \citep{Zubatov1985} & 48 & 48 & 169 & 88 & 40 & $\sim$70\tabularnewline
C-$^{11}$B$_{\textrm{Si}}(k_{\textrm{b}})$ & 0 & 0 & 6 & 2 & 2 & 74\tabularnewline
$^{11}$B-ENDOR \citep{Matsumoto1997} & $-$6.78 & $-$6.78 & 2.40 & $-$3.72 & 3.06 & $\sim$70\tabularnewline
\hline 
$^{13}$C-$\textrm{B}_{\textrm{Si}}(h_{\textrm{a}})$ & 30 & 30 & 182 & 81 & 51 & 0\tabularnewline
$^{13}$C-EPR \citep{Zubatov1985} & 48 & 48 & 173 & 90 & 42 & 0\tabularnewline
C-$^{11}$B$_{\textrm{Si}}(h_{\textrm{a}})$ & 2 & 2 & 8 & 4 & 2 & 0\tabularnewline
$^{11}$B-ENDOR \citep{Matsumoto1997} & $-$3.88 & $-$3.88 & 4.85 & $-$0.97 & 2.91 & 0\tabularnewline
\end{tabular}
\end{ruledtabular}

\end{table}

Regarding the $^{11}$B HF interactions, like their measured analogues,
the amplitudes are very small (few MHz). Unlike the calculations,
the measured Fermi contact is negative. Still, the discrepancy is
well within the estimated error. Hence, along with the $g$ tensors,
the HF calculations provide compelling support for the assignment
of $\textrm{B}_{\textrm{Si}}^{0}$ to the EPR/ENDOR data as reproduced
in Tabs.~\ref{tab1} and \ref{tab2}.

\noindent 
\begin{figure}
\includegraphics[clip,width=8.5cm]{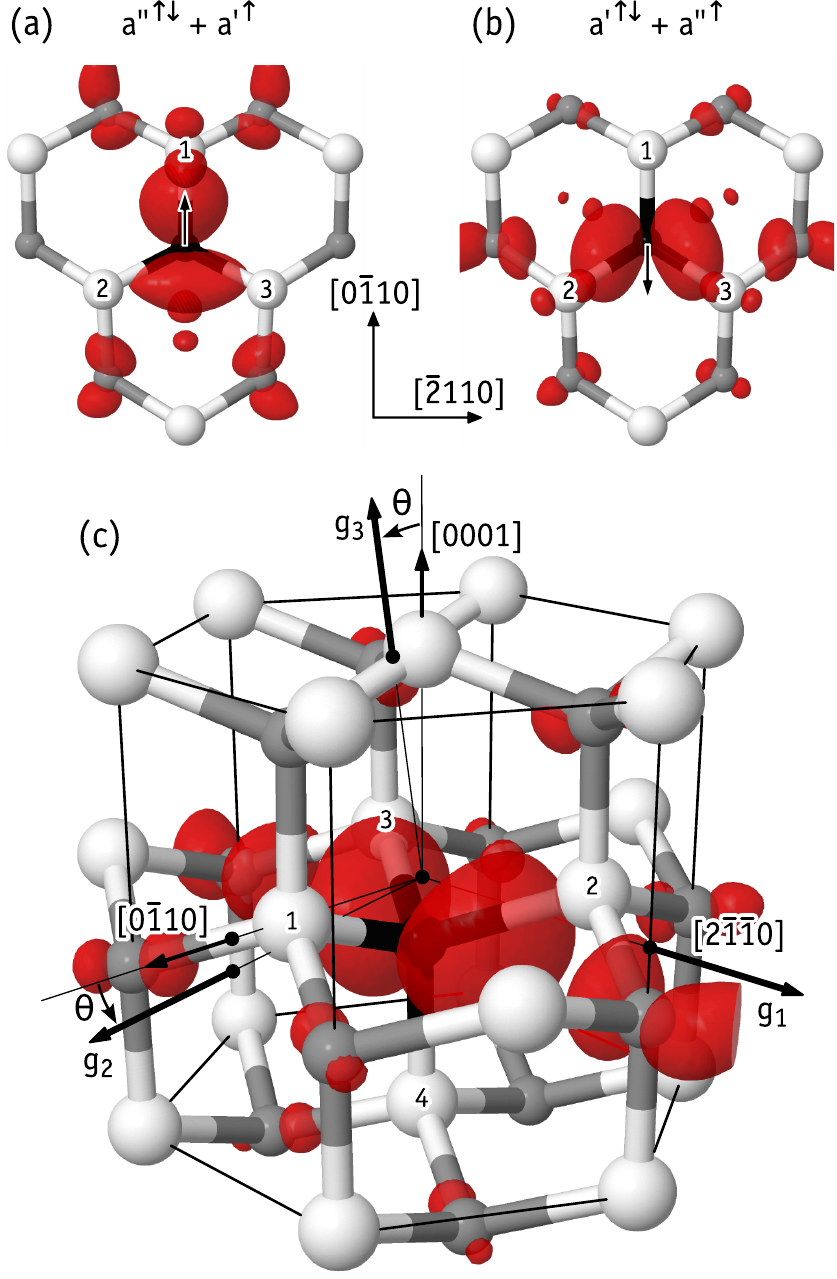}

\caption{\label{fig8}Spin-density isosurface (cutoff 0.02~e/Å$^{3}$) of
a neutral $\textrm{B}_{\textrm{C}}$ defect at the $k$ site of 4H-SiC.
(a) and (b) depict two alternative Jahn-Teller distorted states, namely
$a''^{\uparrow\downarrow}+a'^{\uparrow}$ (metastable) and $a'^{\uparrow\downarrow}+a''^{\uparrow}$
(ground state), which lead to opposite displacements of the B atom
along $[01\bar{1}0]$ (see arrows). In (c) we depict the $a'^{\uparrow\downarrow}+a''^{\uparrow}$
ground state along with the principal directions of the calculated
$g$ tensor (see text). Si, C and B atoms are shown in white, gray
and black, respectively.}
\end{figure}

The above HF interaction calculations were carried out using the GIPAW
code within the GGA to the exchange-correlation potential. We performed
test calculations at the HSE06 level (using the VASP code) and found
that the Fermi contact terms were about a factor of two larger. The
HSE06-level dipolar terms were similar to those found using the semilocal
functional. Such discrepancy was also reported in Ref.~\citep{Skachkov2019}
for the evaluation of isotropic coupling constant using semilocal
and hybrid functionals, and that calls for further investigations.

Regarding the deep boron species, among the arguments behind its assignment
to a $\textrm{B}_{\textrm{Si}}\textrm{-}V_{\textrm{C}}$ structure
were the negligible $^{13}$C and $^{11}$B hyperfine satellites next
to the main signal, as well as a pronounced localization of the spin
density on Si atoms \citep{DuijnArnold1998,Baranov1998}. Unfortunately,
the dynamic Jahn-Teller effect makes any comparison between the measurements
and the static HF calculations rather difficult --- unlike the Zeeman
effect, the $^{29}$Si HF interactions are intermittent due to rotation
of the nodal wave function.

We calculated the $g$ tensor for neutral $\textrm{B}_{\textrm{Si}}(k)\textrm{-}V_{\textrm{C}}^{0}(k)$,
with both the B atom and the vacancy aligned along the crystalline
main axis. The ground state structure involves an electronically-inert
threefold coordinated B atom next to three Si radicals edging the
C-vacancy, two of which reconstruct to form an elongated bond due
to JT effect (see Ref.~\citep{DuijnArnold1998} and references therein).
Most spin density of this complex is localized on a single Si dangling
bond polarized toward the center of the vacancy and the B atom. Although
the distance between B and the Si radical is approximately the separation
between second neighbors of the crystal, Si radical states are rather
extended in space. In fact, considering its symmetry and character,
the paramagnetic state must have a finite amplitude on B atom, and
that feature does not favor the $\textrm{B}_{\textrm{Si}}\textrm{-}V_{\textrm{C}}^{0}$
model.

The calculated $g$ tensor of $\textrm{B}_{\textrm{Si}}\textrm{-}V_{\textrm{C}}^{0}$
allows us draw more definite conclusions. The static JT distorted
state of $\textrm{B}_{\textrm{Si}}(k)\textrm{-}V_{\textrm{C}}^{0}(k)$
with $C_{1h}$ symmetry has a main $g_{3}$ component along the Si
dangling bond, which makes an angle of $\theta=78^{\circ}$ with $[0001]$
-- this was not observed at a temperature as low as $T=4$~K. Accordingly,
two nearly axial EPR signals with a main axis along $[0001]$ were
reported \citep{Baranov1998}. The magnitude of the calculated $g$
values also differ markedly from the observed ones. Even considering
a dynamic JT state (with effective $C_{3v}$ symmetry), the calculated
effective $g$ value of $\textrm{B}_{\textrm{Si}}(k)\textrm{-}V_{\textrm{C}}^{0}(k_{\textrm{dyn}})$
along the $c$ axis ($g_{3}=2.0063$) is too small when compared to
its measured counterpart ($g_{3}=2.029$) \citep{Baranov1998}.

In Sec~\ref{subsec:db} it was shown that the paramagnetic state
of $\textrm{B}_{\textrm{C}}^{0}$ has spin-1/2, and that it derives
from a partially occupied JT distorted $e(xy)^{\uparrow\downarrow\uparrow}$
doublet. Also as detailed on the right hand side of Fig.~\ref{fig2},
this manifold derives from the boron $\textrm{p}_{x}\textrm{p}_{y}$
states, which are nodal on the boron atom as well as along the $c$
axis. The spin density of the JT distorted configurations ${a''}^{\uparrow\downarrow}+{a'}^{\uparrow}$
and ${a'}^{\uparrow\downarrow}+{a''}^{\uparrow}$ is depicted in Figs.~\ref{fig8}(a)
and \ref{fig8}(b). Such a shape anticipates a very small spin localization
on the B atom. For the ${a'}^{\uparrow\downarrow}+{a''}^{\uparrow}$
ground state, the amplitude is zero on boron and high on two basal
Si ligands (Si$_{2}$ and Si$_{3}$).

The spin density of the ground state ${a'}^{\uparrow\downarrow}+{a''}^{\uparrow}$
configuration of $\textrm{B}_{\textrm{C}}^{0}(k)$ is zoomed in Fig.~\ref{fig8}(c).
The case of $\textrm{B}_{\textrm{C}}^{0}(h)$ is analogous and a similar
discussion applies. The figure also depicts the principal directions
of the $g$ values with respect to the crystalline axes. Like it was
considered for the shallow boron defect, trigonal ($C_{3v}$) states
have its main $g_{3}$ component along the $[0001]$ hexagonal axis.
Also, monoclinic ($C_{1h}$) states have $g_{1}$ perpendicular to
the $\{2\bar{1}\bar{1}0\}$ plane, while $g_{2}$ and $g_{3}$ are
rotated by an angle $\theta$ away from $\langle0\bar{1}10\rangle$
and $\langle0001\rangle$, respectively.

Let us first consider the case of static JT distorted configurations.
These correspond to monoclinic states with calculated $g$ values
of $g_{1}\approx2.005$, $g_{2}\approx2.017\textrm{-}2.020$ and $g_{3}\approx2.024\textrm{-}2.028$.
The latter is rotated away from {[}0001{]} by $\theta=13^{\circ}$
only. Although the magnitude of $g_{3}$ is not far from the measured
axial $g$ values, the monoclinic rotation angle was not observed.

Considering that $\textrm{B}_{\textrm{C}}^{0}$ is predicted to show
a dynamic JT effect, the effective $g$ values are better estimated
via averaging over symmetrically equivalent alignments. Hence, we
find $g_{1}=g_{2}\approx2.012$ for both $\textrm{B}_{\textrm{C}}^{0}(k_{\textrm{dyn}})$
and $\textrm{B}_{\textrm{C}}^{0}(h_{\textrm{dyn}})$, whereas $g_{3}\approx2.028$
and $2.024$ for $\textrm{B}_{\textrm{C}}^{0}(k_{\textrm{dyn}})$
and $\textrm{B}_{\textrm{C}}^{0}(h_{\textrm{dyn}})$, respectively.
As reported in Tab.~\ref{tab1}, the calculated main $g_{3}$ values
are in excellent agreement with the axial $g$ values observed for
deep boron defects in 4H-SiC \citep{Baranov1998}. The basal $g$
values also compare well with the corresponding measured figures ($\sim\!2.0$),
although these are accompanied by relatively large error bars due
to random $g$-strain broadening effects \citep{DuijnArnold1998}.

The nodal state shown in Fig.~\ref{fig8}(c) strongly overlaps with
two of the Si atoms connected to boron (Si$_{2}$ and Si$_{3}$).
The two other Si ligands are nodal (Si$_{1}$ and Si$_{4}$) and have
no overlap with the spin density. We suggest that the dynamic JT effect
on this defect could be responsible for an intermittent localization
on all atoms, thus explaining the weak and broad hyperfines detected
for $^{11}$B, $^{13}$C, and $^{29}$Si. Finally, we also note that
the dynamical nature of the ground-state of $\textrm{B}_{\textrm{C}}^{0}$,
and a possible occupancy of both ${a''}^{\uparrow\downarrow}+{a'}^{\uparrow}$
and ${a'}^{\uparrow\downarrow}+{a''}^{\uparrow}$ states above few
tens of degrees Kelvin, could explain the broadening and quenching
of the EPR main signals of deep boron above $T\approx30$~K \citep{Baranov1998}.

\section{Conclusions\label{sec:conclusions}}

We reported on first-principles hybrid density functional calculations
of boron defects in 4H-SiC. Besides defect structures and electronic
transition levels, defect free energies at finite temperatures, $g$
tensor calculations and hyperfine coupling constants were also reported.
The vibrational contribution to the free energies, as well as the
one-electron states for the calculation of the paramagnetic properties,
were found within a semilocal approximation to the electronic exchange
and correlation interactions.

We support the assignment of the shallow boron species to $\textrm{B}_{\textrm{Si}}$.
In the neutral state, these defects possess a threefold coordinated
B atom next to an unsaturated C radical. We mind the reader that this
structure was obtained when the atomistic relaxation was performed
within hybrid DFT. Lower level GGA calculations led to fourfold coordinated
boron atoms. In line with arguments already reported \citep{Gerstmann2004},
the erroneous GGA structure derives from the overmixing between the
acceptor state of boron and the valence band top of the crystal. However,
unlike Ref.~\citep{Gerstmann2004}, we conclude that the neutral
B$_{\textrm{Si}}$ defect adopts a singlet state. The axially distorted
structure of this defect (along the $c$ axis) conserves the maximum
point group symmetry of the 4H-SiC crystal ($C_{3v}$). The displacement
from the perfect lattice site can be explained by the host crystal
field. Hence, the off-site structure cannot be justified by a Jahn-Teller
effect --- it is simply driven by the short covalent radius of boron
compared to that of Si.

As a word of caution, we note that the relative energy of on-site
and off-site B$_{\textrm{Si}}^{0}$ states cannot be easily obtained
with the present method. If the fourfold coordinated B$_{\textrm{Si}}^{0}$
is a diffuse effective-mass-like state, it could be disfavored due
to the artificial confinement effect of the supercell approximation
\citep{Wang2009}. Still, even if that was the case, only the off-site
threefold coordinated B$_{\textrm{Si}}^{0}$ model (and not the EMT-model)
could account for the measurements.

The C radicals on cubic and hexagonal $\textrm{B}_{\textrm{Si}}$
defects are polarized differently, \emph{i.e.}, along basal and axial
bond directions of the crystal, respectively. This feature has been
previously detected by EPR but left unexplained. We demonstrate that
it results from distinct crystal fields acting on each sublattice
site.

Substitutional boron on the carbon site (B$_{\textrm{C}}^{0}$) is
a dynamic Jahn-Teller system with a ``Mexican hat'' like potential.
The potential ripples for rotation around the symmetry axis of the
undisturbed state are 15~meV high only. This figure is lower than
the zero-point energy of the defect, implying that is shows effective
trigonal symmetry, even at liquid-helium temperature.

$\textrm{B}_{\textrm{Si}}$ and $\textrm{B}_{\textrm{C}}$ are both
single acceptors. Despite adopting rather different alignments in
the crystal, the acceptor levels of $\textrm{B}_{\textrm{Si}}(k)$
and $\textrm{B}_{\textrm{Si}}(h)$ are estimated in a narrow range
$E_{\textrm{v}}+(0.34\textrm{-}0.32)$~eV. This could explain the
observation of a single transition by Laplace-DLTS for shallow boron.
The acceptor level of $\textrm{B}_{\textrm{C}}$ is anticipated at
$E_{\textrm{v}}+(0.63\textrm{-}0.67)$~eV, in excellent agreement
with the D-center transition level measured in the range 0.5-0.7~eV
above $E_{\textrm{v}}$. Our results suggest that recently reported
Laplace-DLTS experiments unfolding the D-center signal into two components,
relate to a “shallower” configuration sitting at the cubic carbon
site and a “deeper” one replacing the hexagonal site.

From the calculated free-energies of $\textrm{B}_{\textrm{Si}}$ and
$\textrm{B}_{\textrm{C}}$, we found that under typical growth temperatures,
the equilibrium concentration ratio $[\textrm{B}_{\textrm{Si}}]/[\textrm{B}_{\textrm{C}}]\approx200$
and about $0.1$ for a Si-poor and Si-rich stoichiometry, respectively.
This leads us to the conclusion that formation of $\textrm{B}_{\textrm{C}}$
cannot be avoided during growth when boron is present, and contamination
of n-type layers with boron could limit the mobility and liftetime
of holes due to trapping and recombination at deep $\textrm{B}_{\textrm{C}}$
acceptors.

We demonstrated that the EPR measurements of shallow boron can be
described by a site- and temperature-dependent $g$ tensor of $\textrm{B}_{\textrm{Si}}$.
Below $\sim\!50$~K, neutral $\textrm{B}_{\textrm{Si}}$ defects
at $k$ and $h$ sites show static $C_{1h}$ and $C_{3v}$ symmetry,
with comparable $g$ values along the carbon radical p-state, respectively
$g_{3}=2.0028$ and $2.0022$. These figures compare very well with
2.0025 and 2.0019 from the measurements, respectively. Above $\sim\!50$~K,
the EPR signal related to the hexagonal species remains unchanged.
However, the B-C broken bond in $\textrm{B}_{\textrm{Si}}(k)$ can
reorient by surmounting a barrier of about 0.1~eV, and the estimated
\emph{thermally-averaged} $g_{3}$ value (now parallel to $[0001]$)
increases to 2.0073 (to be compared to 2.0064 from the measurements).

Calculations of the gyromagnetic tensor are complemented with calculations
of the most prominent $^{13}$C and $^{11}$B hyperfine splitting
interactions involving core atoms at the threefold coordinated B$_{\textrm{Si}}^{0}$
defects. The results agree well with the measurements both in terms
of magnitude and axial direction of the interactions.

Our results rule against the assignment of a $\textrm{B}_{\textrm{Si}}\textrm{-}V_{\textrm{C}}$
complex to the deep boron defect. Both directions and magnitude of
the calculated $g$ values for this complex, differ markedly from
the observations. Combining with previous calculations which concluded
that $\textrm{B}_{\textrm{Si}}\textrm{-}V_{\textrm{C}}$ is a donor
without levels in the lower half of the gap \citep{Aradi2001}, we
can definitely abandon the idea of a relation between the deep boron
center and $\textrm{B}_{\textrm{Si}}\textrm{-}V_{\textrm{C}}$.

Instead we assign deep boron to $\textrm{B}_{\textrm{C}}$. The calculated
$g$ values for $\textrm{B}_{\textrm{C}}$ show excellent agreement
with the measurements for deep boron if we account for the dynamics
of the defect. We argue that the dynamic Jahn-Teller effect, along
with the nodal shape of the paramagnetic state, could explain the
weak and broad hyperfine signals related to $^{11}$B, $^{13}$C and
$^{29}$Si. Additionally, by considering $\textrm{B}_{\textrm{C}}$
as being responsible for the deep boron spectra, and hence ruling
out the $\textrm{B}_{\textrm{Si}}\textrm{-}V_{\textrm{C}}$ model,
we naturally avoid having to justify the inexplicable formation of
$\textrm{B}_{\textrm{Si}}\textrm{-}V_{\textrm{C}}$ defects with exclusive
axial orientations as observed by EPR.

\section*{Data availability statement}

The data that support the findings of this study are available from
the corresponding author upon reasonable request.
\begin{acknowledgments}
The present work was supported by the NATO Science for Peace and Security
Programme, project no. G5674. JC and VJBT acknowledge the FCT through
projects LA/P/0037/2020, UIDB/50025/2020 and UIDP/50025/2020.
\end{acknowledgments}

\bibliographystyle{apsrev4-2}

%

\end{document}


\title{Theory of shallow and deep boron defects in 4H-SiC\\
\textmd{\small{}(Supplemental Material to Physical Review B 106, 224112
(2022); \href{https://doi.org/10.1103/PhysRevB.106.224112}{DOI:10.1103/PhysRevB.106.224112})}}
\author{Vitor J. B. Torres}
\affiliation{I3N, Department of Physics, University of Aveiro, Campus Santiago,
3810-193 Aveiro, Portugal}
\author{Ivana Capan}
\affiliation{Ruđer Bošković Institute, Bijenic\'{k}a 54, 10000 Zagreb, Croatia}
\author{José Coutinho}
\affiliation{I3N, Department of Physics, University of Aveiro, Campus Santiago,
3810-193 Aveiro, Portugal}
\email{jose.coutinho@ua.pt}

\maketitle

\section{Convergence of results with the boundary conditions}

First-principles point defect calculations commonly rely on the supercell
approach, implying the use of periodic boundary conditions. Due to
the neutralization condition and limited size of the supercells, the
energy of charged defects is affected by sizable artificial interactions
involving an infinite array of charges trapped at the defects implicitly
present on every cell image, plus that of an explicit or implicit
neutralizing background charge. In order to remove these unwanted
interactions, several methods have been proposed (see for instance
Refs.~{[}50, 51, 54, 56, 69{]}). We use the recipe of Kumagai and
Oba (KO) {[}51{]}, which generalizes the method of Freysoldt, Neugebauer,
and Van de Walle {[}50{]} for anisotropic materials like 4H-SiC.
Previous tests to the KO method, using the double positive carbon
vacancy ($V_{\textrm{C}}^{2+}$) in 4H-SiC as case study, found that
corrected formation energies were affected by an error bar below 20~meV
for cells with few hundred atoms {[}52{]}.

We know however, that the accuracy of such corrections (for a specific
supercell size) depends on the shape and extent of the acceptor/donor
states {[}36, 51, 69{]}. For that reason, we report below a set of
tests pertaining the convergence of the formation energy of B$_{\textrm{Si}}^{-}(h)$
in 4H-SiC using supercells of several sizes. We note that we are dealing
with a singly negative point defect and that the correction varies
with $q^{2}$, where $q$ is the net charge of the defect {[}54{]}.

\noindent 
\begin{figure}
\includegraphics[clip,width=8.5cm]{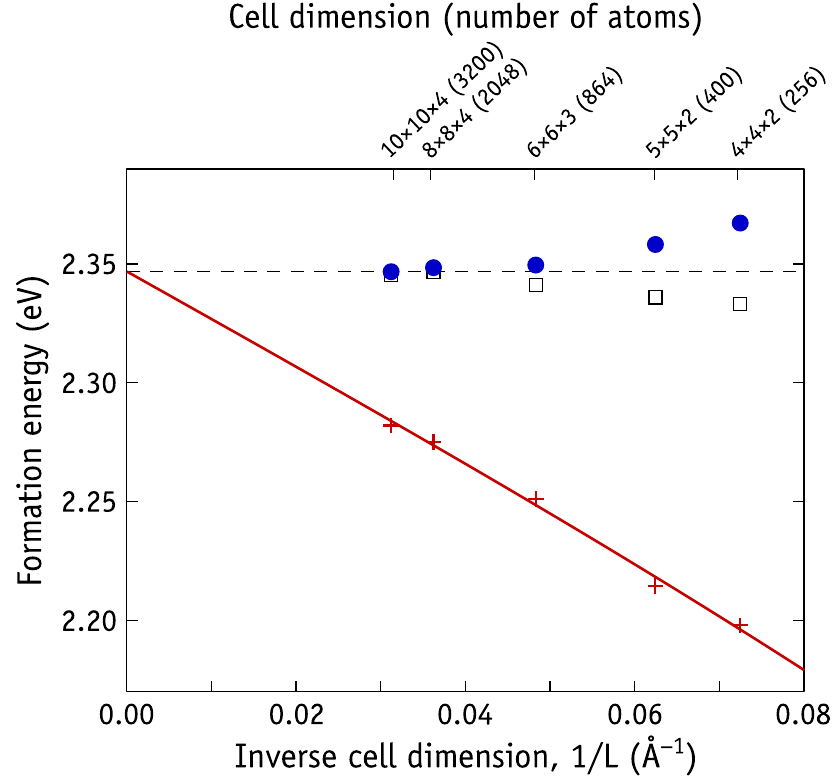}

\caption{\label{fig-s1}Convergence test to the formation energy of $\textrm{B}_{\textrm{Si}}^{-}(h)$
in 4H-SiC. The upper horizontal axis indicates the $m\times m\times n$
unit cell replication factors that correspond to the supercells used,
along with the corresponding total number of atoms within parentheses.
The lower horizontal axis is the inverse of the cell dimension $L^{-1}=V^{-1/3}$,
where $V$ is the volume of the supercell. Crosses represent the bare
(uncorrected) formation energies, open squares are formation energies
corrected by the simplest point charge correction of Makov and Payne
{[}54{]}, and closed circles represent formation energies corrected
with the KO method {[}51{]}.}
\end{figure}

Supercells with the following geometries were employed: $4\times4\times2\,(256)$;
$5\times5\times2\,(400)$; $6\times6\times3\,(864)$; $8\times8\times4\,(2048)$
and $10\times10\times4\,(3200)$. Here, the $l\times m\times n$ triplets
represent the unit cell replication factors and the figure within
parentheses are the corresponding total number of atoms. We define
the effective supercell size as $L=V^{1/3}$, where $V$ is the total
volume. The sampling of the Brillouin zones was carried out over $\Gamma$-centered
$2\!\times\!2\!\times\!2$ and $1\!\times\!1\!\times\!1$ grids for
the two smaller and three larger cells, respectively. Further details
(including the calculation of chemical potentials) are described in
the paper.

Figure~\ref{fig-s1} depicts the results of the convergence tests
to the formation energy of $\textrm{B}_{\textrm{Si}}^{-}(h)$ in C-rich
4H-SiC as a function of $1/L$. The red crosses represent the uncorrected
formation energies. The red line is a fit of a function of the type
$aL^{-1}+bL^{-3}+c$ to the data {[}55{]}. The leading term accounts
for the monopole correction, whereas the term in $L^{-3}$ mostly
accounts for dipole and elastic interactions. The constant $c=2.35$~eV
is the extracted asymptotic formation energy for $L\rightarrow\infty$,
represented by the horizontal dashed line in Fig.~\ref{fig-s1}.
We can readily note that the fit is nearly linear, to large extent
suggesting that the error can be accounted for by a monopole correction.
This is confirmed by the calculated formation energies with a point
charge correction as proposed by Makov and Payne {[}54{]}, shown
in Fig.~\ref{fig-s1} as open squares. For the 400-atom supercells
used in this work the difference to the asymptotic value is about
10~meV. These results are interpreted in the following way: B$_{\textrm{Si}}^{-}$
is a substitutional anion, where the extra electron is mostly localized
on the first neighboring B-C bonds. Hence, it is not surprising that
it can be nearly described as a negative point charge.

Fig.~\ref{fig-s1} also shows the formation energies with the KO
correction as a function of the supercell size (filled circles). The
KO-corrected formation energies slightly overshoot the asymptotic
value. For the 400-atom cells, the point charge and KO corrections
to the negatively charged supercell energies are 0.12~eV and 0.14~eV,
respectively. From the above, we estimate that the periodic boundary
conditions induce an error bar of $\sim\!20$~meV to the calculated
(KO-corrected) formation energies.

\section{Defect geometries\label{sec:geometries}}

In Tab.~\ref{tab-s1} we list the first and second neighbor distances
to the B atom in substitutional boron defects in 4H-SiC. PBE relaxations
were carried out with a $\Gamma$-centered $2\times2\times2$ sampling
mesh, whereas $\Gamma$ sampling was done for the HSE06 relaxations.

\noindent 
\begin{table}
\begin{ruledtabular}
\caption{\label{tab-s1}Calculated first (B-$X_{1}$) and second (B-$X_{2}$)
neighbor distances to the B atom in substitutional boron defects replacing
Si and C sites of 4H-SiC. Results using both PBE and HSE06 functionals
are listed. All distances are in Å.}

\noindent %
\begin{tabular}{lllll}
Defect & \multicolumn{2}{c}{B-$X_{1}$} & \multicolumn{2}{c}{B-$X_{2}$}\tabularnewline
Functional & PBE & HSE06 & PBE & HSE06\tabularnewline
\hline 
$\textrm{B}_{\textrm{Si}}^{0}(k)$ & 1.73 & 1.65 & 1.75 & 2.41\tabularnewline
$\textrm{B}_{\textrm{Si}}^{0}(h)$ & 1.75 & 1.65 & 1.75 & 2.43\tabularnewline
$\textrm{B}_{\textrm{Si}}^{-}(k)$ & 1.75 & 1.76 & 1.75 & 1.76\tabularnewline
$\textrm{B}_{\textrm{Si}}^{-}(h)$ & 1.75 & 1.76 & 1.76 & 1.77\tabularnewline
\hline 
$\textrm{B}_{\textrm{C}}^{0}(k)$ & 1.92 & 1.92 & 1.95 & 1.98\tabularnewline
$\textrm{B}_{\textrm{C}}^{0}(h)$ & 1.92 & 1.92 & 1.94 & 1.97\tabularnewline
$\textrm{B}_{\textrm{C}}^{-}(k)$ & 1.90 & 1.90 & 1.90 & 1.91\tabularnewline
$\textrm{B}_{\textrm{C}}^{-}(h)$ & 1.90 & 1.90 & 1.91 & 1.91\tabularnewline
\end{tabular}
\end{ruledtabular}

\end{table}

In Ref.~{[}36{]} its was found that defect structures obtained using
$\Gamma$ sampling on 216-atom cells, could be sensitive upon increasing
the density of the mesh. Ref.~{[}47{]} reports that for the case
of neutral carbon interstitial in 3C-SiC, the PBE relaxed structure
(and spin state) were incorrect due to mixing of defect states with
the conduction band.

To investigate the reliability of the $\Gamma$-sampled 400-atom structures
obtained within HSE06, we can inspect the atomistic geometries for
the GIPAW calculations (HF splittings and $g$ tensor), which were
evaluated within HSE06 in 256-atom supercells using $\Gamma$-centered
$2\times2\times2$ $\mathbf{k}$ point grids (see Sec.~II of the
paper). The resulting structures and relative energies were essentially
identical to those found from the $\Gamma$-sampled calculations.
For instance, first and second B-C neighboring distances for the offsite
B$_{\textrm{Si}}^{0}(h)$ ground state were 1.65 and 2.45~Å. For
B$_{\textrm{Si}}^{0}(k)$, the analogous distances are 1.65 and 2.42~Å.
These figures are very close to the structural parameters reported
in Tab.~\ref{tab-s1}.

We also carried out a batch of tests where we started from the B$_{\textrm{Si}}^{0}(h)$
ground state structure obtained in 400-atom cells using $\Gamma$
sampling and HSE06, and performed a PBE-level relaxation using different
$\mathbf{k}$ point meshes. The tests considered $\Gamma$-centered
$2\times2\times2$ and $4\times4\times4$, as well as Monkhorst and
Pack ($\Gamma$-free) $2\times2\times2$ and $4\times4\times4$ grids
{[}68{]}. All these calculations provide the same incorrect answer,
$i.e.$, the final structure is a fourfold coordinated B$_{\textrm{Si}}$
defect without a carbon radical.

\section{Reorientation of neutral B$_{\mathbf{Si}}$}

\noindent 
\begin{figure}
\includegraphics[clip,width=8.5cm]{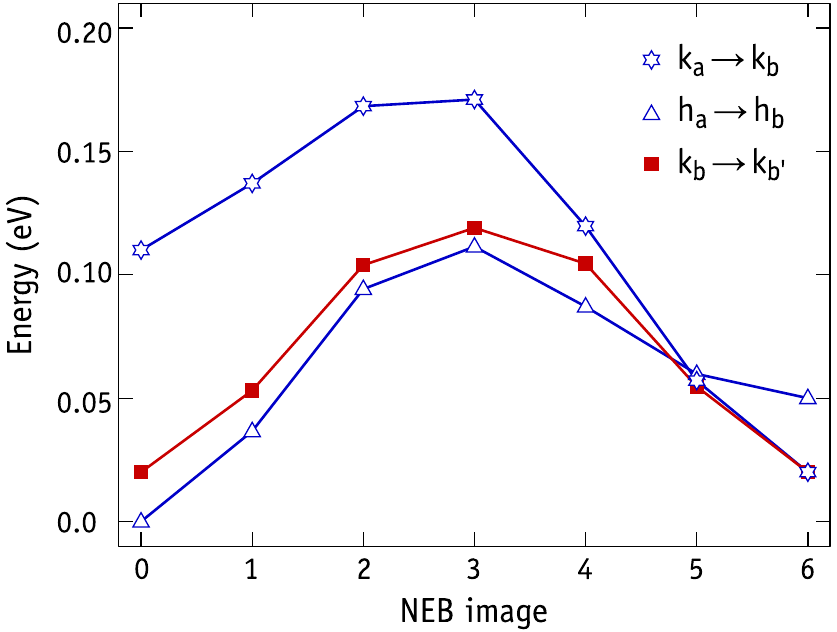}

\caption{\label{fig-s2}Minimum energy paths for boron motion in B$_{\textrm{Si}}^{0}$
between the stable/metastable structures indicated in the legend.
Calculations were performed using the climbing image nudged elastic
band method {[}75{]}. Open symbols correspond to reorientations mechanisms
from axial to basal offsite configurations. Filled squares correspond
to the reorientation of B$_{\textrm{Si}}^{0}(k)$ between two neighboring
basal distortions. All mechanisms consider 5 intermediate images between
the endpoints.}
\end{figure}

Here we provide the details behind the calculation of the barriers
considered in the configuration coordinate diagram of Fig.~1(c) of
the paper. They pertain to the reorientation of $\textrm{B}_{\textrm{Si}}^{0}$
between offsite configurations in both $k$ and $h$ sublattice sites
($k_{\textrm{a}}\rightarrow k_{\textrm{b}}$ and $h_{\textrm{a}}\rightarrow h_{\textrm{b}}$).
We also detail the calculation of the minimum energy path for the
reorientation of $\textrm{B}_{\textrm{Si}}^{0}(k)$ between two neighboring
(symmetry equivalent) offaxis basal configurations ($k_{\textrm{b}}\rightarrow k_{\textrm{b}'}$).
This figure is relevant for understanding the temperature dependence
of the electron paramagnetic resonance data (see Sec.~III.E of the
paper).

The reorientation mechanisms were investigated using the climbing
image nudged elastic band method (CI-NEB) {[}75{]}. Essentially,
we started with a guessed sequence by setting up an array of 400-atom
supercell structures, comprising five intermediate structures linearly
interpolated between the stable end-configurations. On a second step,
a CI-NEB relaxation was performed with forces calculated at the HSE06
level and using the $\Gamma$-point for sampling the BZ. In a final
step, the energies of the relaxed sequence of images were further
refined by improving the $\mathbf{k}$ point sampling to $\Gamma$-centered
$2\!\times\!2\!\times\!2$ and performing single-point energy calculations
at the HSE06 level.

Figure~\ref{fig-s2} shows the result of the NEB calculations referred
above. At the ends (horizontal axis coordinates 0 and 6) we find ground
states for each sublattice location ($k_{\textrm{b}}$, $h_{\textrm{a}}$
and $k_{\textrm{b}'}$) and metastable states ($k_{\textrm{a}}$ and
$h_{\textrm{b}}$) of offsite B$_{\textrm{Si}}^{0}$ defects. All
energies are represented with respect to the lowest energy B$_{\textrm{Si}}^{0}(h_{\textrm{a}})$
state. From the data we find activation barriers of 0.06~eV and 0.11~eV
for $k_{\textrm{a}}\rightarrow k_{\textrm{b}}$ and $h_{\textrm{a}}\rightarrow h_{\textrm{b}}$
reorientations, respectively. A reorientation barrier of 0.09~eV
is found for the $k_{\textrm{b}}\rightarrow k_{\textrm{b}'}$ jump
of B$_{\textrm{Si}}^{0}(k)$.

\section{Overmixing of the shallow boron acceptor with the valence band within
PBE}

\begin{figure}
\includegraphics[clip,width=8.5cm]{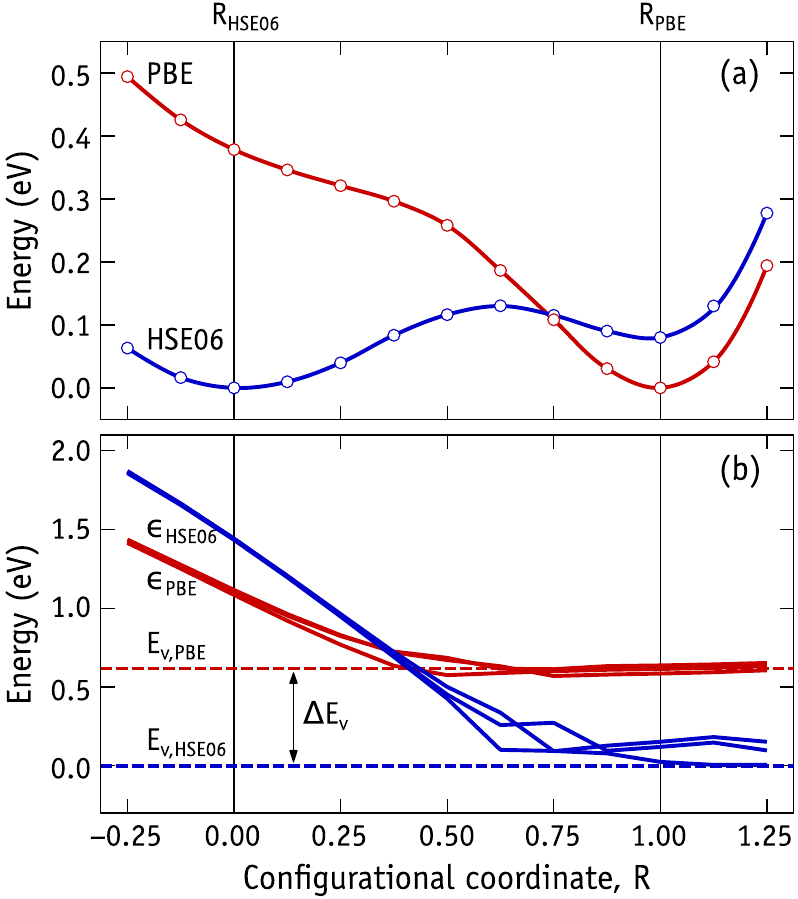}

\caption{\label{fig-s3}Analysis of neutral $\textrm{B}_{\textrm{Si}}$ in
4H-SiC as a function of a configurational coordinate $R$ (see text),
using semilocal (PBE) and hybrid (HSE06) density functional theory.
(a) Total energies of $\textrm{B}_{\textrm{Si}}^{0}(h)$ defects at
several $R$ coordinates using PBE (red) and HSE06 (blue) functionals.
The origin of the energy scale of each plot corresponds to the respective
minimum energy structure. (b) Lowest unoccupied Kohn-Sham states of
$\textrm{B}_{\textrm{Si}}^{0}(h)$ in 4H-SiC using PBE and HSE06 functionals
($\epsilon_{\textrm{PBE}}$ and $\epsilon_{\textrm{HSE06}},$respectively).
A total of three Kohn-Sham states are represented calculation, corresponding
to three different $\mathbf{k}$ points. The dashed lines represent
the valence band top energy as found from the highest occupied Kohn-Sham
state in a bulk supercell (using the appropriate functional). All
energies are relative to $E_{\textrm{v,HSE06}}$. PBE data (red) is
shifted from the HSE06 data (blue) by a valence band offset $\Delta E_{\textrm{v}}=0.63$~eV
(see text).}
\end{figure}

Here we justify our argument regarding the existence of an overestimated
and spurious mixing effect found at the PBE level, between the lowest
unoccupied state of B$_{\textrm{Si}}^{0}$ with the valence band of
4H-SiC. In Fig.~\ref{fig-s3}(a) we depict two sequences of single-point
total energies of B$_{\textrm{Si}}^{0}(h)$ defects in 4H-SiC. They
are both represented as a function of a reaction coordinate $R$,
where atomic coordinates $\mathbf{R}=\mathbf{R}_{\textrm{HSE06}}(1-R)+\mathbf{R}_{\textrm{PBE}}R$
are linearly interpolated between ground state coordinates of B$_{\textrm{Si}}^{0}(h)$
at HSE06 ($R=0$, offsite structure) and PBE level (onsite structure).
The calculations were spin polarized, used 400-atom supercells and
$\Gamma$-centered $2\!\times\!2\!\times\!2$ grids for the BZ sampling.
For each functional, the origin of the energy scale is set to the
respective ground state structure.

\noindent 

As expected, the red line in Fig.~\ref{fig-s3}(a) joining the sequence
of PBE energies, shows a clear minimum at $R=1$ ($\mathbf{R}_{\textrm{PBE}}$)
while the offsite $\mathbf{R}_{\textrm{HSE06}}$ configuration is
not stable. This was confirmed upon structural relaxation of the $\mathbf{R}_{\textrm{HSE06}}$
coordinates within PBE. In Sec.~\ref{sec:geometries} of this document,
we shown that this is not an artifact and does not change upon improving
the $\mathbf{k}$ point sampling.

Conversely, at the HSE06 level (blue line), the minimum energy is
at $R=0$ ($\mathbf{R}_{\textrm{HSE06}}$) and the $\mathbf{R}_{\textrm{PBE}}$
structure is metastable by nearly 0.1~eV. A relaxation of the onsite
$\mathbf{R}_{\textrm{PBE}}$ structure at the HSE06 level resulted
in a local minimum (also four fold coordinated boron) 0.06~eV above
the relaxed ground state $R=0$.

Let us now look at the band structure change along the coordinate
$R$. This is shown in Fig.~\ref{fig-s3}(b). Line colors and labels
are in direct correspondence with Fig.~\ref{fig-s3}(a). We note
that the valence band maxima of 4H-SiC within PBE lies $\Delta E_{\textrm{v}}=0.63$~eV
above that of HSE06 (both shown as horizontal dashed lines). We are
using the same psudopotentials in the semilocal and hybrid calculations.
Under these conditions, the macroscopically averaged PBE- and HSE06-level
electrostatic potentials across the bulk are almost identical. Hence,
$\Delta E_{\textrm{v}}$ could be trivially found from the difference
in the valence band maxima eigenvalues {[}78{]}. This operation,
although not strictly necessary, has been carried out for the sake
of clarity of the figure.

The energy of the lowest unoccupied state of $\textrm{B}_{\textrm{Si}}^{0}(h)$
is represented along $R$ by the solid lines. For each calculation,
a total of three energies are represent and they stand for eigenvalues
at three different $\mathbf{k}$ points. This gives us an idea of
the band dispersion for this level as a function of defect geometry.
Importantly, for all structures, the highest occupied state (spin-up
channel) has the character of the valence band and its localization
spans across the supercell. Any spin-up B-related occupied state must
be located below $E_{\textrm{v}}$.

It is clear that for both functionals, the lowest unoccupied state
is high in the gap for the offsite $\mathbf{R}_{\textrm{HSE06}}$
configuration. On the other hand, inspection of the highest occupied
and lowest unoccupied states for the $\mathbf{R}_{\textrm{PBE}}$
configuration (within both PBE and HSE06) confirms that they are both
delocalized and resemble a valence band state. In other words, as
we go from threefold to fourfold coordination of boron, the unoccupied
deep gap state of the offsite configuration \emph{submerges} into
the valence band, becoming occupied and leaving a hole at the top
of the valence band.

Like for any gap state, the HSE approximation also mixes the boron
state with the band edges. That is particularly evident for the $\mathbf{R}_{\textrm{PBE}}$
structure, where a fourth B-C bond is formed. However, while the relevant
gap state of the $\mathbf{R}_{\textrm{HSE06}}$ geometry is nearly
1.5~eV above $E_{\textrm{v}}$ within HSE-level DFT (this is about
mid gap), within PBE the same structure shows a gap state which is
less than 0.5~eV above the corresponding $E_{\textrm{v}}$ level.
This allows for less mixing in the correct HSE result, and more band
mixing in the flawed PBE calculation.

While PBE suggests that B$_{\textrm{Si}}$ in 4H-SiC is an effective
mass theory (EMT) like acceptor, HSE06 results point for a deep hole
trap (with a level about 0.3~eV above $E_{\textrm{v}}$). These conclusions
are also in line with the staggering difference between the spin density
of threefold and fourfold coordinated $\textrm{B}_{\textrm{Si}}^{0}(h)$
defects (see Figs~3(a) and 3(b) of the paper). Unlike the PBE diffuse
spin density, within HSE06 the density is strongly localized on the
nearest shells of atoms around B, most notably on the C radical next
to boron.

\section{Hole capture mechanism of B$_{\mathbf{Si}}$}

\noindent 
\begin{figure}
\includegraphics[clip,width=8.5cm]{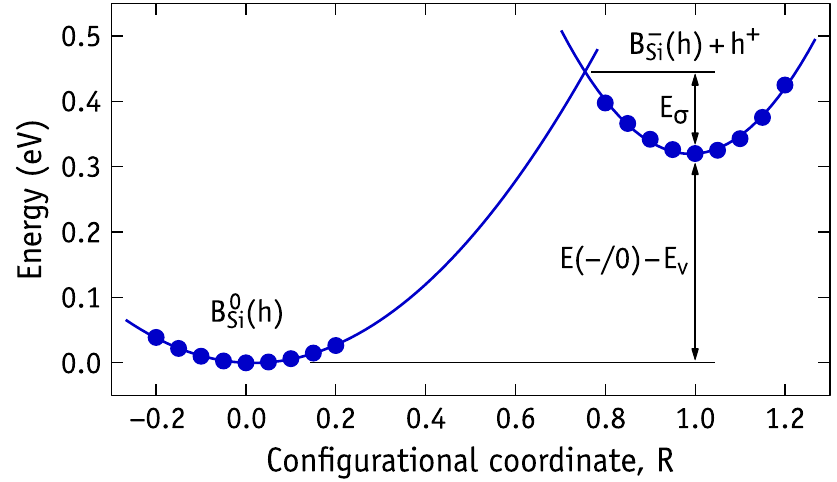}

\caption{\label{fig-s4}Potential energy surface along a linear path joining
ground state structures of B$_{\textrm{Si}}^{0}(h)$ (with $R=0$)
and B$_{\textrm{Si}}^{-}(h)$ (with $R=1$). Solid lines represent
parabolic fits to the data in the vicinity of the respective minima.}
\end{figure}

The $\textrm{B}_{\textrm{Si}}^{-}$ defect does not have deep states
in the gap --- it is a fourfold coordinated isovalent anion whose
negative charge is localized on its four B-C bonds. It is possible
that this state could attract free holes and hold them with a small
EMT like binding energy. Indeed as indicated in Fig.~\ref{fig-s3},
the onsite structure of B$_{\textrm{Si}}^{0}$ is metastable. Our
method is not the best to estimate meV-accurate effective mass hole
binding energies. That would require many thousands of atoms in the
supercell {[}79{]}. However, judging from the experimental and calculated
spectroscopic properties of B$_{\textrm{Si}}$, the neutral ground
state is the offsite threefold coordinated structure. An eventual
onsite $\textrm{B}_{\textrm{Si}}^{0}$ defect would be an excited
state. We identify two possible mechanisms for hole capture: (1) A
two step mechanism involving hole capture in the onsite negative state,
followed by reconfiguration to the neutral ground state, $\textrm{B}_{\textrm{Si}}^{-}(\textrm{on})+h^{+}\rightarrow\textrm{B}_{\textrm{Si}}^{0}(\textrm{on})\rightarrow\textrm{B}_{\textrm{Si}}^{0}(\textrm{off})$;
(2) Direct hole capture involving a strong electron phonon coupling,
$\textrm{B}_{\textrm{Si}}^{-}(\textrm{on})+h^{+}\rightarrow\textrm{B}_{\textrm{Si}}^{0}(\textrm{off})$.
Here, `on/off' stand for on/offsite configurations.

From Fig.~\ref{fig-s3}(a) the barrier for the second step of mechanism
(1) is estimated as $\lesssim0.1$~eV. As for mechanism (2), the
capture of the hole by negatively charged $\textrm{B}_{\textrm{Si}}^{-}(h)$
must involve a concomitant local deformation of the B-C bonds, so
that the \emph{moving} level shown in Fig.~\ref{fig-s3}(b), which
in this case is occupied, will emerge into the gap. A simple estimate
of that barrier was found by calculating the potential energy of the
defect for structures in the vicinity of offsite and onsite ground
states of neutral and negatively charged B$_{\textrm{Si}}(h)$. These
correspond to atomic coordinates $\mathbf{R}^{0}$ and $\mathbf{R}^{-}$,
respectively. Again, we define a one-dimensional coordinate $R$,
such that all nuclear coordinates are constrained along an effective
coupling direction of motion $\mathbf{R}=\mathbf{R}^{0}(1-R)+\mathbf{R}^{-}R$
(see for instance Refs.~{[}80, 81{]} for further details).

Figure~\ref{fig-s4} shows the potential energy surface along the
linear path joining ground state structures of B$_{\textrm{Si}}^{0}(h)$
(with $R=0$) and B$_{\textrm{Si}}^{-}(h)$ (with $R=1$). The energy
of the two minima are separated in the energy scale by the calculated
depth of the hole trap, $E(-/0)-E_{\textrm{v}}=0.32$~eV. A parabolic
fit to the potential data in the vicinity of the two minima allows
us to estimate the hole capture barrier for $\textrm{B}_{\textrm{Si}}^{-}(h)+h^{+}\rightarrow\textrm{B}_{\textrm{Si}}^{0}(h)$
as about $E_{\sigma}\sim0.1$~eV. Two additional features in Fig.~\ref{fig-s4}
are highlighted --- (i) the two parabola are not nested, suggesting
that the capture mechanism involves a strong electron-phonon coupling,
typical of deep centers {[}77{]}; (ii) the effective mode frequency
of the neutral state (proportional to the curvature of the lower parabola)
is substantially softer than that of the negative state. This is consistent
with the formation of a B-C bond upon hole emission (or electron capture)
by $\textrm{B}_{\textrm{Si}}^{0}$.

It is clear that the EPR data, in particular the $^{13}$C hyperfine
splitting for this center {[}26, 29{]}, cannot be accounted for by
an EMT like onsite $\textrm{B}_{\textrm{Si}}$ acceptor. On the contrary,
the bistable model described above, proposes that $\textrm{B}_{\textrm{Si}}$
is a deep hole trap, offering a physical explanation for the observations.